\documentclass[11pt]{tep}
\usepackage[a4paper,margin=2.5cm]{geometry}
\pdfoutput=1 
\usepackage{www}

\begin{document}
\thispagestyle{empty}
\def\thefootnote{\fnsymbol{footnote}}
\setcounter{footnote}{1}
\null
\mbox{}\hfill  
FR-PHENO-2019-007, TIF-UNIMI-2019-12
\vskip0cm
\vfill
\begin{center}
  {\Large \boldmath{\bf 
      Next-to-leading-order QCD and electroweak corrections to\\
      triple-$\PW$ production with leptonic decays at the LHC
    }
    \par} \vskip2.5em
  {\large
    {\sc Stefan Dittmaier$^{1}$, Gernot Knippen$^{1}$ 
      and Christopher Schwan$^{2}$
    }\\[1ex]
    {\normalsize \it 
      $^1$ Albert-Ludwigs-Universit\"at Freiburg, Physikalisches Institut, \\
      79104 Freiburg, Germany \\
      $^2$ Tif Lab, Dipartimento di Fisica, Università di Milano and INFN,
      Sezione di Milano, \\
      20133 Milano, Italy
    }
    \\[2ex]
  }
  \par \vskip1em
\end{center}\par
\vskip.0cm \vfill {\bf Abstract:} \par
We present a calculation of the next-to-leading-order QCD and electroweak corrections to $\PW\PW\PW$ production with leptonically decaying \PW bosons at the LHC, fully taking into account off-shell contributions, intermediary resonances, and spin correlations.
The contributions of the quark--antiquark-induced electroweak correction to typical fiducial cross sections at the LHC are of the order of 5--8\,\% and grow to tens of percent in the high-energy tails of distributions.
We observe strong cancellations among the positive quark--photon and negative quark--antiquark-induced electroweak corrections.
In addition to results based on full $2\to 6/7$-particle next-to-leading-order matrix elements, we present a calculation based on the triple-pole approximation, which expands the matrix elements around the poles of three simultaneously resonant \PW bosons.
The triple-pole approximation performs particularly well for integrated cross sections and for differential cross sections that are insensitive to off-shell effects, such as angular and rapidity distributions.
\par
\vskip3cm
\noindent
December 2019
\par
\null
\setcounter{page}{0}
\clearpage
\def\thefootnote{\arabic{footnote}}
\setcounter{footnote}{0}
\section{Introduction}
For a detailed understanding of fundamental interactions in nature, precision tests of the Standard Model (SM) are mandatory.
In particular, we want to deepen our knowledge of electroweak symmetry breaking (EWSB).
To gain such knowledge, it is of great importance to pursue Higgs precision physics and to investigate and measure multi-boson processes like $\mathrm{WWW}$ production.
Precise predictions for integrated and differential cross sections are needed to confront theory with data and to obtain possible constraints on physics beyond the SM (BSM), which might manifest itself in anomalous triple or quartic gauge couplings if described by an effective field theory.
The production of three \PW bosons in proton--proton collisions is one of the few processes that provide the possibility to constrain the quartic $\mathrm{WWWW}$ coupling directly and is therefore of special interest.
Many BSM models modify the EWSB as realized in the SM which further motivates exploring multi-boson production processes, because those processes are very sensitive to on- and off-shell Higgs-boson exchange.
There is ongoing effort in observing triple-\PW boson production at the LHC \cite{Aaboud:2016ftt,CMS:2019mpq}, and recently evidence was established \cite{Aad:2019udh}.

The QCD corrections to $\Pp\Pp\rightarrow\PW\PW\PW+X$ with \cite{Campanario:2008yg} and without \cite{Binoth:2008kt} leptonic decays have been known for more than ten years.
Additionally, results matched to parton showers were presented in Ref.~\cite{Hoeche:2014rya}.
Next-to-leading order (NLO) electroweak (EW) corrections together with NLO QCD corrections in an improved narrow-width approximation and for on-shell \PW bosons were calculated in Refs.~\cite{Yong-Bai:2016sal} and \cite{Dittmaier:2017bnh}, respectively.
In particular, large positive contributions from quark--photon-induced channels that cancel the negative quark--antiquark-induced corrections were observed, rendering the precise knowledge of the photon PDF~\cite{Manohar:2016nzj,Manohar:2017eqh,Harland-Lang:2019eai} particularly important.
In Ref.~\cite{Frederix:2018nkq} the NLO EW corrections to on-shell $\mathrm{WWW}$ production were worked out as well.
More recently, the NLO EW corrections with full off-shell \PW bosons were presented in Ref.~\cite{Schonherr:2018jva}.
In this article we provide an independent check of the off-shell results based on full $2\rightarrow 6/7$-particle amplitudes and combine the EW with the QCD corrections.
While Ref.~\cite{Schonherr:2018jva} used \propername{Recola} 1.2 \cite{Actis:2012qn,Actis:2016mpe} as one-loop matrix element provider only, we have performed two independent calculations which employ \propername{OpenLoops} 2 \cite{Cascioli:2011va,Kallweit:2014xda,Buccioni:2019sur} and \propername{Recola} 1.4, respectively.

Furthermore, we present a comparison of the full off-shell calculation with a calculation done within the triple-pole approximation (TPA), which is based on the leading pole term in a threefold resonance expansion.
To our knowledge, this is the first time a pole approximation is being used for three resonances.
The construction of the TPA generalizes in a straightforward way the concept of a double-pole approximation to describe \PW-pair production with leptonic decays at the LHC \cite{Billoni:2013aba}, which was used before for \PW-pair production in $\Pem\Pep$ collisions \cite{Denner:1999gp,Denner:2000bj,Denner:2002cg}\footnote{Alternative forms of the double-pole approximation were presented in Refs.~\cite{Beenakker:1998gr,Jadach:1998tz}, and a comparison of the different approximations in Ref.~\cite{Grunewald:2000ju}.}.
In particular, we further extend the comparison of pole approximation and full off-shell calculation, which was presented for \PW-pair production in $\Pem\Pep$ \cite{Denner:2005es,Denner:2005fg} and $\Pp\Pp$ collisions \cite{Biedermann:2016guo} before.

The paper is structured in the following way:
We describe the partonic processes and the ingredients of the NLO calculation in Sec.~\ref{sec:process}, followed by a description of the TPA in Sec.~\ref{sec:tpa}.
In Sec.~\ref{sec:results}, we present numerical results of our calculation.
In detail, we describe the input-parameter scheme used in the numerical calculations in Sec.~\ref{sec:results_parameters} and subsequently present numerical results of the full off-shell calculation in Secs.~\ref{sec:results_offshellint} and \ref{sec:results_offshelldiff}.
In Sec.~\ref{sec:results_tpa-vs-full}, we discuss results obtained in the TPA and compare them with the full off-shell calculation.
Finally, in Sec.~\ref{sec:Conclusion}, we conclude with a summary.

\section{Triple-\PW production at proton--proton colliders}
\label{sec:process}
We consider the two charge-conjugated processes
\begin{equation}
  \Pp\Pp\rightarrow\Pem\Pane\Pmp\Pnm\Ptp\Pnt+X \qquad \text{and} \qquad \Pp\Pp\rightarrow\Pep\Pne\Pmm\Panm\Ptm\Pant+X
  \label{eq:process_partonic}
\end{equation}
with three different lepton generations.
At leading order (LO) these processes are induced by the partonic subprocesses
\begin{equation}
  u_i\bar{d}_j\rightarrow\Pem\Pane\Pmp\Pnm\Ptp\Pnt \qquad \text{and} \qquad \bar{u}_id_j\rightarrow\Pep\Pne\Pmm\Panm\Ptm\Pant,
\end{equation}
respectively, where $i$ and $j$ indicate the fermion generations.
We neglect mixing with the third quark generation and do not consider the top quark as a parton of the proton at LHC energies.
Therefore, the participating quarks are \Pqu, \Pqd, \Pqs, \Pqc, and their antiquarks.
In the production of the three \PW bosons, already at LO triple and quartic gauge vertices occur.
Moreover, the final state contains associated production of a Higgs boson together with a \PW boson.

As long as all leptons are considered massless, the cross section for the first process of Eq.~\eqref{eq:process_partonic} is, up to negligible interference terms, equal to the cross section for $\Pp\Pp\rightarrow\Pem\Pane\Pmp\Pnm\Pmp\Pnm+X$ and $\Pp\Pp\rightarrow\Pmm\Panm\Pep\Pne\Pep\Pne+X$ after multiplication with the correct symmetry factor of $\nicefrac{2}{4}$.
The situation is analogously valid for the charge-conjugated case, i.e.\ the second process of Eq.~\eqref{eq:process_partonic} and $\Pp\Pp\rightarrow\Pep\Pne\Pmm\Panm\Pmm\Panm+X$ or $\Pp\Pp\rightarrow\Pmp\Pnm\Pem\Pane\Pem\Pane+X$.

\begin{figure}
\centering
\begin{subfigure}[b]{0.33\textwidth}
\centering
\includegraphics{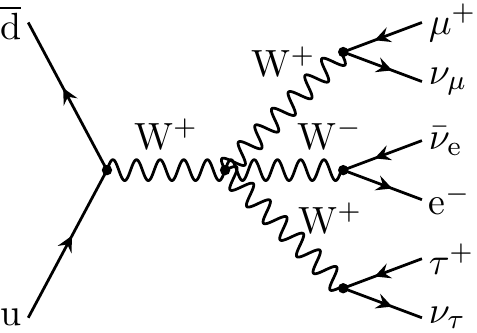}
\caption{$s$-channel with 3 resonances}
\label{fig:triply-res-s-channel}
\end{subfigure}%
\begin{subfigure}[b]{0.33\textwidth}
\centering
\includegraphics{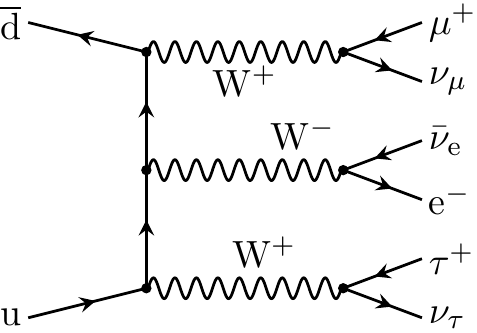}
\caption{$t$-channel with 3 resonances}
\label{fig:triply-res-t-channel}
\end{subfigure}%
\begin{subfigure}[b]{0.33\textwidth}
\centering
\includegraphics{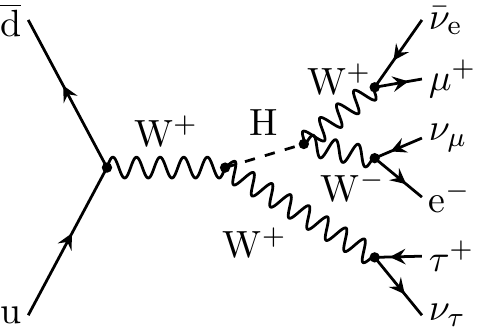}
\caption{associated Higgs production}
\label{fig:associated-higgs-production}
\end{subfigure}%
\par\bigskip
\begin{subfigure}[b]{0.33\textwidth}
\centering
\includegraphics{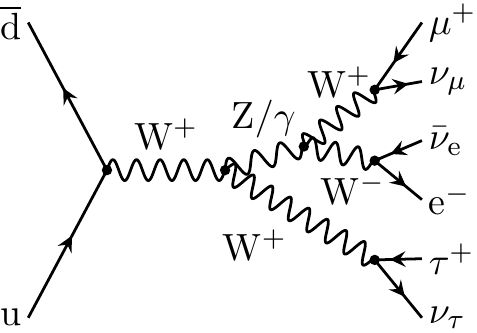}
\caption{$\PW\PZ$ production}
\label{fig:higgs-replaced-by-z-photon}
\end{subfigure}%
\begin{subfigure}[b]{0.33\textwidth}
\centering
\includegraphics{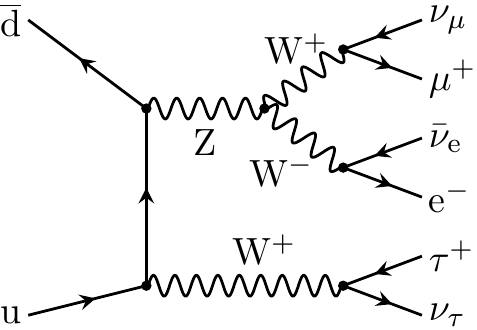}
\caption{single $t$-channel}
\label{fig:single-t-channel}
\end{subfigure}%
\begin{subfigure}[b]{0.33\textwidth}
\centering
\includegraphics{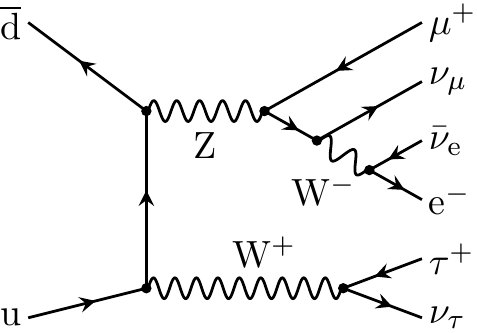}
\caption{sequential resonances}
\label{fig:resonance-chain}
\end{subfigure}%
\caption{Examples of LO Feynman diagrams contributing to $\Pp\Pp\rightarrow\Pem\Pane\Pmp\Pnm\Ptp\Pnt+X$.}
\label{fig:lo-diagrams}
\end{figure}

At LO, there are three basic classes of diagrams that involve up to three resonances and contribute to the cross section of the $\mathrm{WWW}$ production process $\Pp\Pp\rightarrow\Pem\Pane\Pmp\Pnm\Ptp\Pnt+X$ and the corresponding charged-conjugated process:
\begin{enumerate}
\item Diagrams with three simultaneously resonant \PW bosons (e.g.\ Fig.~\ref{fig:triply-res-s-channel}--\ref{fig:single-t-channel}),
\item Higgs production in association with a \PW boson (e.g.\ Fig.~\ref{fig:associated-higgs-production}), where the produced Higgs boson further decays into an on- and an off-shell \PW boson, and
\item $\mathrm{WZ}$ production, where the \PZ boson either decays into an on- and an off-shell \PW boson (e.g.\ Fig.~\ref{fig:higgs-replaced-by-z-photon}) or into a four-fermion state via a resonant \PW~boson (e.g.\ Fig.~\ref{fig:resonance-chain}).
\end{enumerate}
All other diagrams show less resonance enhancement.
The production of $\mathrm{WZ}$ is strongly suppressed because of the four-body decay of the \PZ boson, while associated Higgs production and triply-resonant $\PW\PW\PW$ contributions dominate the cross sections of the given processes.
Due to the extremely narrow width of the Higgs boson and the fact that the Higgs-boson mass is smaller than twice the \PW-boson mass, associated Higgs production is well separated from the triply-resonant $\PW\PW\PW$ contributions in phase space and therefore can be isolated by phase-space cuts.

\begin{figure}
\centering
\begin{subfigure}[b]{0.33\textwidth}
\centering
\includegraphics{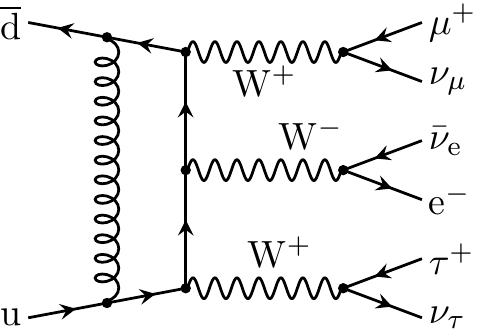}
\caption{QCD pentagon loop}
\label{fig:nlo-qcd-virtual}
\end{subfigure}%
\begin{subfigure}[b]{0.33\textwidth}
\centering
\includegraphics{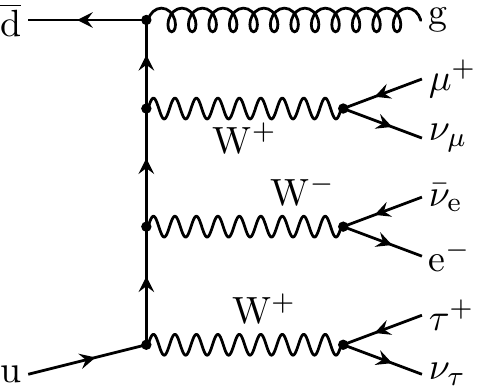}
\caption{gluon radiation}
\label{fig:nlo-qcd-real}
\end{subfigure}%
\begin{subfigure}[b]{0.33\textwidth}
\includegraphics{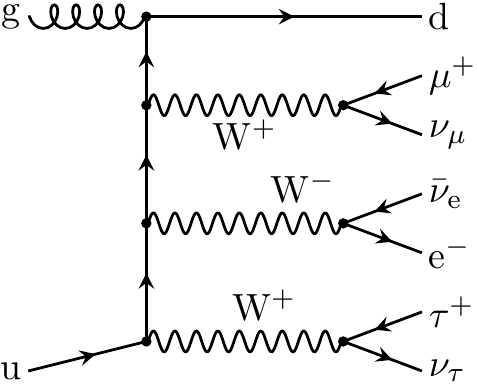}
\caption{quark--gluon induced}
\label{fig:quark-gluon-diagram}
\end{subfigure}%
\caption{Selection of NLO QCD Feynman diagrams contributing to $\Pp\Pp\rightarrow\Pem\Pane\Pmp\Pnm\Ptp\Pnt+X$.}
\label{fig:nlo-qcd-diagrams-plus-real-ew}
\end{figure}

In Fig.~\ref{fig:nlo-qcd-virtual} we show a loop diagram contributing to the NLO QCD correction, and in Fig.~\ref{fig:nlo-qcd-real} a corresponding real emission diagram.
Figure \ref{fig:quark-gluon-diagram} shows a diagram for the quark--gluon-induced real correction.
\begin{figure}
\centering
\begin{subfigure}[b]{0.33\textwidth}
\centering
\includegraphics{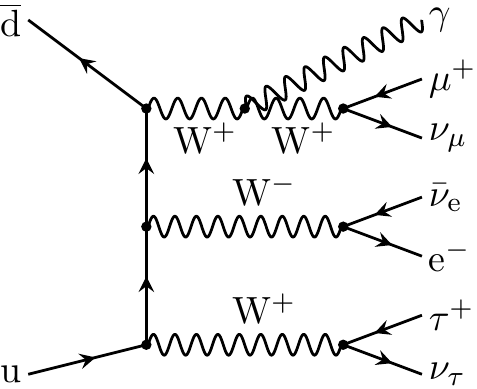}
\caption{photon radiation}
\label{fig:nlo-ew-real}
\end{subfigure}%
\begin{subfigure}[b]{0.33\textwidth}
\includegraphics{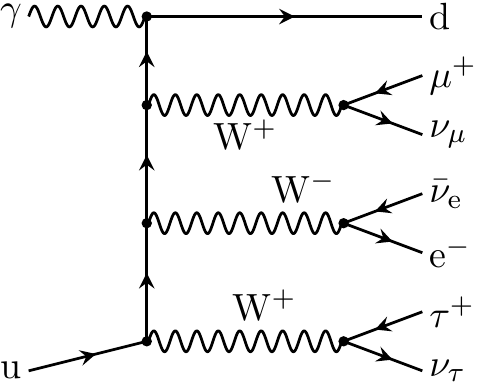}
\caption{quark--photon induced}
\label{fig:quark-photon-diagram}
\end{subfigure}%
\caption{Examples for NLO EW real emission diagrams contributing to $\Pp\Pp\rightarrow\Pem\Pane\Pmp\Pnm\Ptp\Pnt+X$.}
\label{fig:real-ew-diagrams}
\end{figure}
Figure \ref{fig:real-ew-diagrams} depicts two diagrams of NLO EW real emission.
Note that quark--photon-induced contributions, as shown in Fig.~\ref{fig:quark-photon-diagram}, are the only contributions at NLO EW with an additional jet.
\begin{figure}
\centering
\begin{subfigure}[b]{0.33\textwidth}
\centering
\includegraphics{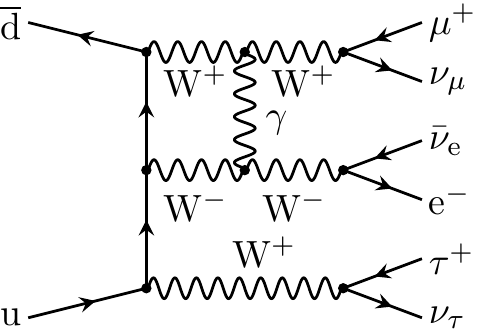}
\caption{non-factorizable type mm'}
\label{fig:nlo-ew-mmprime}
\end{subfigure}%
\begin{subfigure}[b]{0.33\textwidth}
\centering
\includegraphics{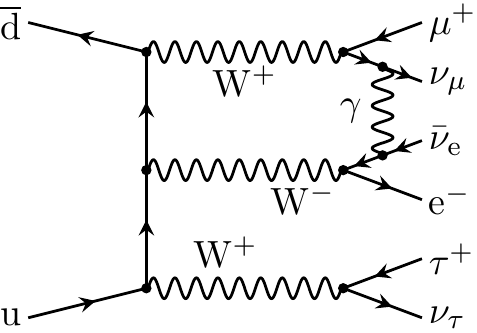}
\caption{non-factorizable type ff'}
\label{fig:nlo-ew-ffprime}
\end{subfigure}%
\begin{subfigure}[b]{0.33\textwidth}
\centering
\includegraphics{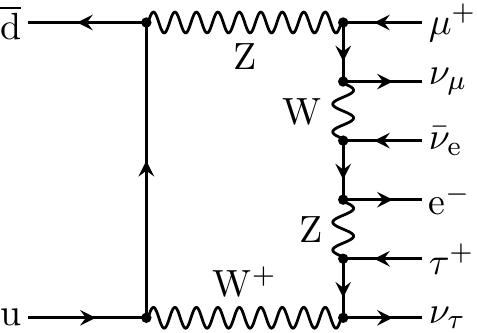}
\caption{8-point function}
\label{fig:nlo-ew-8pt-fun}
\end{subfigure}%
\caption{Examples of virtual NLO EW Feynman diagrams contributing to $\Pp\Pp\rightarrow\Pem\Pane\Pmp\Pnm\Ptp\Pnt+X$.}
\label{fig:virtual-nlo-ew-diagrams}
\end{figure}
Some NLO EW virtual diagrams are illustrated in Fig.~\ref{fig:virtual-nlo-ew-diagrams}.

We calculate the virtual corrections in two different ways: Firstly, we use full $2\to 6$-particle amplitudes of the off-shell process, and secondly, we evaluate the virtual correction in TPA.

We have implemented two fully independent calculations for all ingredients, in particular for the virtual one-loop contributions, for the real emission parts, and for the multidimensional phase-space integration.
The off-shell calculation and the TPA are carried out as follows:
One calculation uses amplitudes provided by \propername{OpenLoops 2} \cite{Cascioli:2011va,Kallweit:2014xda,Buccioni:2019sur} and, in the case of the pole approximation, amplitudes generated by in-house software based on Feynman diagrams generated with \propername{FeynArts 1} \cite{Kublbeck:1990xc}.
The second implementation uses amplitudes generated with \propername{MadGraph} \cite{Alwall:2014hca} and \propername{Recola} \cite{Actis:2012qn,Actis:2016mpe}, and for the pole-approximated virtual corrections amplitudes created by \propername{FeynArts} \cite{Hahn:2000kx} and \propername{FormCalc} \cite{Hahn:1998yk}, which are further processed and modified.
Tensor and scalar loop integrals are evaluated using the \propername{Collier} library \cite{Denner:2016kdg}, which uses the techniques and results described in Refs.~\cite{Denner:2002ii,Denner:2005nn,Denner:2010tr} and supports internal complex masses for unstable particles as required by the complex-mass scheme \cite{Denner:1999gp,Denner:2005fg,Denner:2006ic}.
Both implementations use adaptive multi-channel Monte Carlo integration techniques \cite{Hilgart:1992xu,Kleiss:1994qy} with independent implementations of Feynman-diagram-inspired phase-space mappings for the individual channels.

The real corrections as well as the LO cross sections are always calculated with full $2\to 7/6$ matrix elements.
We subtract IR singularities, which arise due to soft and/or collinear emission of gluons, photons, or additional quarks, using the dipole subtraction formalism \cite{Catani:1996vz,Dittmaier:1999mb,Catani:2002hc,Dittmaier:2008md}.

We have compared our calculation against the existing NLO QCD results in the literature \cite{Campanario:2008yg} in Tab.~\ref{tab:comparison_campanario} and found agreement.
We could, however, not reproduce the NLO EW results of Ref.~\cite{Schonherr:2018jva}, see Tab.~\ref{tab:comparison_schoenherr}.
In order to clarify the differences to the results of Ref.~\cite{Schonherr:2018jva}, we have contacted the author and compared the individual numerical contributions from the LO, virtual EW and real EW contributions.
In the course of this tuned comparison it turned out that in the calculation of Ref.~\cite{Schonherr:2018jva}, an inconsistent scale choice was used and the cut setup was not exactly as described.
An update of the results of Ref.~\cite{Schonherr:2018jva} shows good agreement with our findings\footnote{Marek Schönherr, private communications.}.
\begin{table}
  \centering
  \caption{Comparison of LO and NLO QCD cross sections, \slo and \snloqcd, of the process $\Pp\Pp\rightarrow\Pem\Pane\Pmp\Pnm\Ptp\Pnt+X$ with parameters as given by Ref.\ \cite{Campanario:2008yg} and a Higgs mass of $120\,\GeV$.
    In contrast to Ref.\ \cite{Campanario:2008yg}, which uses a modified complex-mass scheme with a real mixing angle $\theta_W$,  we employ the usual complex-mass scheme where the weak mixing angle is a complex quantity (see Eq.~\eqref{eq:parameters_mixingangle}).
    Monte Carlo integration errors are indicated in parentheses after each result.}
  \label{tab:comparison_campanario}
  \begin{tabular}{cd{1.8}d{1.8}}
    \toprule
    Ref. & \ccol{\slo {\small [\fb{}]}} & \ccol {\small\snloqcd [\fb{}]}\\
    \midrule
    \cite{Campanario:2008yg} & 0.2256(2) & 0.3589(4)\\
    our calculation & 0.22548(5) & 0.35944(14)\\
    \bottomrule
  \end{tabular}
\end{table}
\begin{table}
  \centering
  \caption{Comparison of the LO cross section \slo and relative NLO EW corrections $\delta_{q\bar{q}}^\EW$, $\delta_{q\gamma/\bar{q}\gamma}^\EW$ of the process $\Pp\Pp\rightarrow\Pem\Pane\Pmp\Pnm\Pmp\Pnm+X$ with parameters and definitions as given by Ref.\ \cite{Schonherr:2018jva}.
    Monte Carlo integration errors are indicated in parentheses after each result.}
  \label{tab:comparison_schoenherr}
  \begin{tabular}{cd{1.9}d{1.3}d{1.7}}
    \toprule
    Ref. & \ccol{\slo {\small [\fb{}]}} & \ccol{$\delta_{q\bar{q}}^\EW$ {\small [\%]}} & \ccol{$\delta_{q\gamma/\bar{q}\gamma}^\EW$ {\small [\%]}}\\
    \midrule
    \cite{Schonherr:2018jva} & 0.0955 & -4.6 & 2.4 \\
    our calculation & 0.095480(11) & -8.5(4) & 2.3906(11)\\
    \bottomrule
  \end{tabular}
\end{table}
\section{Triple-pole approximation}
\label{sec:tpa}
\subsection{The basic concept}
Similarly to the double-pole approximation for \PW-pair production, the triply-resonant \PW-boson region can be well described by a triple-pole approximation (TPA).
In the TPA we expand transition matrix elements around the three \PW resonances.
In detail, we formulate the TPA similar to the double-pole approximation that was suggested with the Monte Carlo program \propername{RacoonWW} for $\Pep\Pem\rightarrow\PW\PW\rightarrow 4f(+\gamma)$ \cite{Denner:1999gp,Denner:2000bj,Denner:2002cg} which was later also applied to $\PW\PW$ production in proton--proton collisions \cite{Billoni:2013aba}.
As the intrinsic relative precision of the TPA is \order{\nicefrac{\GW}{\MW}} at LO, it is clear that, in order to reach a precision at or below the percent level, the TPA should only be applied to NLO corrections, while LO contributions should be evaluated fully off shell.
We follow the approach to only calculate the virtual correction in the TPA while calculating the LO cross section and the real corrections fully off shell.
Beside the TPA, which is only valid where the three simultaneously resonant \PW bosons dominate the cross section, we alternatively evaluate the virtual correction off shell based on full $2\to 6$ amplitudes.

\begin{figure}
  \centering
  \begin{tikzpicture}
  [baseline={([yshift=-0.95ex]current bounding box.center)},
  line width=1pt,
  phot/.style={decorate,decoration={snake, segment length=#1, amplitude=2pt}},
  phot/.default=9pt,
  ferm/.style={postaction={decorate},decoration={markings, mark=at position #1 with {\stealtharrow}}},
  ferm/.default=0.5,
  higgs/.style=dashed,
  scale=0.5]
  \footnotesize
  \coordinate (i1) at (-1,2);
  \coordinate (i2) at (-1,-2);
  \node[circle,draw,minimum width=20,thick,fill=gray!40] (cprod) at (1.5,0) {};
  \node[circle,draw,minimum width=15,thick,fill=gray!40] (cdec1) at (4.5,2) {};
  \node[circle,draw,minimum width=15,thick,fill=gray!40] (cdec2) at (4.5,0) {};
  \node[circle,draw,minimum width=15,thick,fill=gray!40] (cdec3) at (4.5,-2) {};
  \coordinate (o1) at (7,2.5);
  \coordinate (o2) at (7,1.5);
  \coordinate (o3) at (7,0.5);
  \coordinate (o4) at (7,-0.5);
  \coordinate (o5) at (7,-1.5);
  \coordinate (o6) at (7,-2.5);
  \draw[ferm] (i2) -- (cprod);
  \draw[ferm] (cprod) -- (i1);
  \draw[phot] (cprod) -- node[pos=0.6,above=2pt] {\PW} (cdec1);
  \draw[phot] (cprod) -- node[pos=0.6,above] {\PW} (cdec2);
  \draw[phot] (cprod) -- node[pos=0.6,above] {\PW} (cdec3);
  \draw[ferm] (o1) -- (cdec1);
  \draw[ferm] (cdec1) -- (o2);
  \draw[ferm] (o3) -- (cdec2);
  \draw[ferm] (cdec2) -- (o4);
  \draw[ferm] (o5) -- (cdec3);
  \draw[ferm] (cdec3) -- (o6);
\end{tikzpicture}
  \caption{Structure of diagrams contributing to the factorizable NLO virtual corrections in the TPA. The gray circles indicate either a tree-like substructure or a loop subdiagram.}
  \label{fig:tpa_fact_structure}
\end{figure}
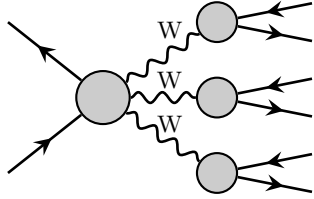
In the pole-approximated virtual correction, two independent gauge-invariant contributions arise: the factorizable and the non-factorizable contributions.
The former consist of all contributions in which the loop corrections can be attributed to either production or decay of the \PW bosons (see Fig.~\ref{fig:tpa_fact_structure}), while the latter comprises the particle exchanges between the production and the decay subprocesses.
An example for a diagram contributing to both factorizable and non-factorizable corrections is given Fig.~\ref{fig:nlo-ew-mmprime}, while Fig.~\ref{fig:nlo-ew-ffprime}, in which a photon is exchanged between two charged leptons from different resonances, contributes to the non-factorizable corrections only.
Figure \ref{fig:nlo-ew-8pt-fun} is an example of a diagram that is neither included in the factorizable nor in the non-factorizable corrections of the TPA and only appears in the full off-shell calculation.
Only soft photon exchange contributes to the non-factorizable correction in the leading pole approximation which leads to a factorization into a correction factor $\delta_\nfac$ and the pole-approximated LO matrix element $\cM_{\LO,\PA}$, so that
\begin{equation}
  2\Re\bigl(\cM_{\LO,\PA}^*\cM_{\text{virt},\text{non-fact.},\PA}\bigr) = \delta_\nfac \absq{\cM_{\LO,\PA}}.
\end{equation}
In the construction of the non-factorizable correction, we follow Ref.~\cite{Dittmaier:2015bfe}, where the non-fac\-tor\-i\-za\-ble corrections for pair production processes \cite{Melnikov:1995fx,Beenakker:1997ir,Denner:1997ia,Denner:2000bj,Accomando:2004de} were generalized to any number of resonances.

\subsection{On-shell projection}

Since matrix elements for the factorizable corrections of the TPA employ separate matrix elements for the production and the decay subprocesses, gauge invariance demands that the momenta of the resonances defining the expansion points are on shell.
This requires an \emph{on-shell projection} of the off-shell phase space.
Such projections are not completely determined by the on-shell requirement, and different projections will yield results which differ within the uncertainty of the pole approximation.

We use an on-shell projection,
\begin{equation}
\{ p_1, p_2, p_3 \} \quad \longmapsto \quad \{ \hat{p}_1, \hat{p}_2, \hat{p}_3 \},
\end{equation}
which simultaneously projects all three momenta $p_i$ of the \PW resonances on shell, so that $\hat{p}_i^2 = M_\PW^2$.
Specifically, our chosen variant of the projection retains the direction of the momenta of two resonances and---if possible---the energy of one of those.
Given the momenta $p_i=(p_i^0,\vec p_i)$ of the three intermediary resonances in their CM frame defined by $\vec p_1+\vec p_2+\vec p_3=0$, a CM energy $\sqrt{s}$, and angle $\alpha_{12}$ between $\vec p_1$ and $\vec p_2$, we define the following shorthands
\begin{equation}
  \begin{aligned}
    \Upsilon &\equiv \sqrt{s}\bigl(s+m_1^2-m_2^2-m_3^2\bigr), & \Delta_\alpha &\equiv s-m_2^2\sin^2\alpha_{12}, \\
    \Delta &\equiv 2\sqrt{s}\bigg(\frac{\sqrt{s}}{2}-\hat p_1^0\bigg)+m_1^2, \quad &\Delta_{23} &\equiv m_2^2-m_3^2,
  \end{aligned}
\end{equation}
as well as the quantities
\begin{align}
    p_{1,+}^0 &\equiv \frac{\Upsilon}{2\Delta_\alpha}\Biggl\{1-\sqrt{1-\frac{\Delta_\alpha}{\Upsilon^2}\left(\left(s+m_1^2+\Delta_{23}\right)^2-4m_2^2(s+m_1^2\cos^2\alpha_{12})\right)}\Biggr\},\\
  p_{1,-}^0 &\equiv p_{1,+}^0\Big\rvert_{\cos\alpha_{12}=0}=\frac{\Upsilon}{2(s-m_2^2)}\Biggl\{1-\sqrt{1-\frac{s-m_2^2}{\Upsilon^2}\left(\left(s+m_1^2+\Delta_{23}\right)^2-4m_2^2s\right)}\Biggr\},\\
    p_{1,b}^0 &\equiv\frac{\sqrt{s}}{2} + \frac{m_1^2-(m_2+m_3)^2}{2 \sqrt{s}}.
\end{align}
In the same frame, we project the energy $p_1^0$ of the first resonance to
\begin{equation}
  \hat{p}_1^0 = \max\Bigl(m_1 + \Delta m, \min\bigl(p_{1,\pm}^0-\Delta m, p_1^0\bigr)\Bigr),
\end{equation}
where the upper limit $p_{1,\pm}^0$ depends on the sign of $\cos\alpha_{12}$ and is $p_{1,+}^0$ for $\cos\alpha_{12}\geq 0$ and $p_{1,-}^0$ for $\cos\alpha_{12} < 0$.
The technical parameter $\Delta m$ is introduced to avoid the kinematical limits of phase space and should be chosen small, i.e.\ $\Delta m\ll m_1$.
A convenient value for $\Delta m$ was found to be
\begin{equation}
  \Delta m = \min\biggl(10^{-3}\,\GeV,\frac{p_{1,b}^0-m_1}{3}\biggr).
\end{equation}
The energy $\hat p_2^0$ of the second resonance after the on-shell projection is given by
\begin{equation}
  \begin{aligned}
    \hat p_2^0=\frac{1}{2\left(\Delta+\vec{\hat{p}}_1^2\sin^2\alpha_{12}\right)}\Bigg[
    &\left(\sqrt{s}-\hat{p}_1^0\right)\left(\Delta+\Delta_{23}\right)\\[-5pt]
    &\pm\sqrt{\vec{\hat{p}}_1^2\cos^2\alpha_{12}\left(\left(\Delta+\Delta_{23}\right)^2-4m_2^2\big(\Delta+\vec{\hat{p}}_1^2\sin^2\alpha_{12}\big)\right)}\Bigg],
  \end{aligned}
\end{equation}
and the spatial parts $\mathbf{\hat{p}}_i$ of the momenta $i=1,2$ read
\begin{equation}
  \mathbf{\hat{p}}_i = \sqrt{\left(\hat{p}_i^0\right)^2-m_i^2}\frac{\vec{p}_i}{\abs{\vec{p}_i}}.
\end{equation}
The on-shell-projected momentum of the third resonance is given by momentum conservation.
We chose the permutation $\pi$ assigning the momenta of the \PW resonances $p_\PWm$, $p_{\PW_1^+}$ ,$p_{\PW_2^+}$ to $p_i$, $i=1,2,3$ with
\begin{equation}
  \{p_1,p_2,p_3\}=\pi\Bigl\{p_\PWm,p_{\PW_1^+},p_{\PW_2^+}\Bigr\}
\end{equation}
in such a way that---if possible---$\hat p_1^0=p_1^0$ and $\lvert \hat p_2^0-p_2^0 \rvert$ ist minimal, resulting in only slightly deformed momenta in most cases.
Having calculated the on-shell-projected momenta of the resonances, the momenta of the external particles are computed using the projections for the $1 \to 2$ decays as described in Ref.~\cite{Dittmaier:2015bfe}, preserving the directions of the charged decay leptons in the CM frame of the respective resonance.

As an alternative way to project the general phase space on shell, we use a \emph{sequential pairwise} on-shell projection as presented in Ref.~\cite{Dittmaier:2015bfe}.
In our case, for the process $\Pp \Pp \to \PWm \PW^+_1 \PW^+_2$, we first project the pair ($\PWm, \PW^+_1$), then the pair ($\PW^+_1, \PW^+_2$).
For the construction of the final-state momenta we chose to preserve the direction of the charged leptons.

We have compared the simultaneous with the sequential pairwise on-shell projection and found a difference of $0.43\,\%$ relative to LO in integrated NLO cross sections based on the two different projection variants.
This difference is within the expected uncertainty of the TPA.
In differential cross sections, differences up to $\sim1.5\,\%$ can be observed.
On-shell projections can by construction only be performed above the production threshold and generally start to break down already in the vicinity.
As the pole approximation is only valid several $\Gamma_\PW$ above the production threshold (see below) this is unproblematic.
Nevertheless, the simultaneous on-shell projection has the advantage over the pairwise on-shell projects that it exists even closely above the $\PW\PW\PW$ production threshold where the pairwise on-shell projection already ceases to be valid.

\subsection{Off-shell Coulomb singularity}
Whenever a pair of on-shell \PW bosons becomes non-relativistic, a Coulomb singularity builds up due to long-range photon exchange between the slowly moving \PW bosons.
In those regions the NLO EW correction effectively behaves as
\begin{equation}
  \delta_\text{Coul} \sim \pm\frac{\alpha\pi}{2\beta_\PW},
\end{equation}
where $\beta_\PW$ is the \PW-boson velocity in the $\mathrm{WW}$ rest frame.
Including instability effects of the \PW bosons, the $\nicefrac{1}{\beta_\PW}$ singularity is regularized by the finite width of the \PW boson.
As the precise form of the off-shell singularity is known \cite{Fadin:1993kg,Bardin:1993mc,Fadin:1994pm}, it is convenient to include the full off-shell Coulomb singularity in the TPA.
The off-shell effects of the Coulomb singularity are already partially accounted for in the non-factorizable correction.
To fully include the off-shell effects we subtract the Coulomb singularity for all pairs of on-shell \PW bosons in the TPA and restore the full off-shell Coulomb singularity by adding \cite{Denner:1997ia}
\begin{equation}
  \begin{aligned}
    \Delta_\text{Coul}=\pm \frac{\alpha}{\pi} \Re\Bigg[&
    2\pi \mathrm{i} \frac{2\MW^2-s_{ij}}{\bar{\beta}s_{ij}}
    \ln\biggl(\frac{\beta+\Delta_M-\bar{\beta}}{\beta+\Delta_M+\bar{\beta}}\biggr)\\
    &- 2\pi \mathrm{i} \frac{2\MW^2-\hat{s}_{ij}}{\beta_\PW\hat{s}_{ij}}
    \ln\biggl(\frac{K_i+K_j+\beta_\PW\Delta_M s_{ij}}{2\beta_W^2\hat{s}_{ij}}\biggr)
    \Bigg]
  \end{aligned}
\end{equation}
for each individual pair $i$, $j$ of \PW resonances to the non-factorizable correction $\delta_\nfac$.
The sign depends on the charges of the two \PW bosons, i.e.\ $-$ for like-sign and $+$ for opposite-sign intermediary \PW-boson pairs.
The invariants $s_{ij}$ and $\hat{s}_{ij}$ are the squared CM energies of the off-shell and the on-shell-projected \PW pair, respectively.
The inverse off-shell propagators $K_{i/j}$ of the resonances read
\begin{equation}
  K_{i/j}=p_{i/j}^2-\mu_W^2,
\end{equation}
with complex squared \PW-boson mass $\mu_\PW^2$ defined in Eq.~\eqref{eq:parameters_complexmass} and resonance four-momentum $p_i^\mu$, which is given by the sum of the two respective decay momenta.
The parameters $\beta_\PW$ and $\bar{\beta}$ are the velocities of the on-shell and off-shell \PW bosons,
\begin{equation}
  \beta_\PW =\sqrt{1-\frac{4\MW^2}{\hat{s}_{ij}}+\mathrm{i} \epsilon},\qquad\bar{\beta}=\frac{\sqrt{\lambda\bigl(s_{ij},p_i^2,p_j^2\bigr)}}{s_{ij}},
\end{equation}
with the Källén function $\lambda(a,b,c)=(a-b-c)^2-4bc$.
We have further introduced the shorthands
\begin{equation}
  \beta=\sqrt{1-\frac{4\mu_\PW^2}{s_{ij}}},\qquad
  \Delta_M=\frac{\bigl\lvert p_i^2-p_j^2 \bigr\rvert}{s_{ij}}.
\end{equation}
Including the full off-shell Coulomb singularity within the TPA calculation changes the integrated cross sections by $\sim0.5\%$ at current LHC center-of-mass energies.

\subsection{Differences between the TPA and the full off-shell $2\to 6/7$ calculation}
The technical advantage of the TPA is that on-shell $\PW\PW\PW$ production is a much simpler process in comparison to the full off-shell process:
In the TPA, we only have to evaluate loop amplitudes of the $2\rightarrow 3$ production and the $1\rightarrow 2$ decay processes with real masses in the internal non-resonant propagators.
For the non-factorizable contributions, generic results are known in a process-independent form, so that they are easy to evaluate and do not complicate the calculation further.
On the other hand, for the full off-shell calculation loop diagrams with up to 8-point functions (e.g.\ Fig.~\ref{fig:nlo-ew-8pt-fun}) have to be evaluated.
This difference in complexity is naturally reflected in the time needed for the numerical evaluation of the loop amplitudes.
For example, comparing the evaluation time for a single phase-space point of the EW one-loop off-shell amplitude, provided by \propername{Recola} 1.4, with one of our TPA amplitudes we observe that the off-shell amplitude needs roughly 7 times longer.

In the case of the off-shell calculation, we work in the complex-mass scheme \cite{Denner:1999gp,Denner:2005fg,Denner:2006ic} where the squared mass $\mu_i^2$ of particle $i$ is complex and given by
\begin{equation}
  \mu_i^2=M_i^2-i\Gamma_iM_i,\quad i=\PW,\PZ,\PH.
  \label{eq:parameters_complexmass}
\end{equation}
In the complex-mass scheme the weak mixing angle also becomes a complex quantity,
\begin{equation}
  \cos\theta_W = \frac{\mu_\PW}{\mu_\PZ},
  \label{eq:parameters_mixingangle}
\end{equation}
to ensure the gauge independence of the loop amplitudes.
The complex-mass scheme guarantees NLO accuracy both in resonant and non-resonant regions of phase space.
In the TPA, real masses are used in the amplitudes for production and decays, complex masses are only used in the resonance propagators.

In general, we expect the TPA to be a good approximation to the full off-shell matrix elements in regions of phase space where all three \PW bosons can become simultaneously resonant.
One caveat of the processes analyzed here, however, is that there is a large contribution to the integrated cross section coming from doubly-resonant $\PW\PH$ production.
Nevertheless, due to the extremely small width of the Higgs boson and the mass hierarchy $\MH<2\MW$ it is possible to exclude this phase-space region and to consider it separately using already existing results for $\PW\PH$ production \cite{Denner:2011id,deFlorian:2016spz,Granata:2017iod} and subsequent $\PH\to\PW\PW\to2\Pl 2\Pn$ decay \cite{Bredenstein:2006rh}.
A quantitative analysis of the approximate quality of the TPA, excluding the Higgs-strahlung contribution, is presented in Sec.~\ref{sec:results_tpa-vs-full}.

It is important to note that the TPA is only valid for partonic scattering energies $\sqrt{\hat s}$ several \GW above the production threshold for three \PW bosons at $\sqrt{\hat s}=3\,\MW$.
Near the threshold region the loop corrections involve the additional small energy scale $\sqrt{\hat s}-3\MW$ beside \MW\ and $\sqrt{\hat s}$.
Taking the TPA in this region would therefore result in a degradation of the TPA accuracy by some factor of $\order{\frac{\GW}{\sqrt{\hat s}-3\MW}}\gtrsim 1$.
Owing to the large suppression of the cross-section contributions near and below the $\PW\PW\PW$ threshold, we can neglect the virtual corrections for $\sqrt{\hat s}<3\MW+10\,\GeV$ and base the TPA predictions on LO and real corrections in this region only.

\section{Numerical results}
\label{sec:results}
\subsection{Input parameters}
\label{sec:results_parameters}
In the following, we use the latest values of the physical on-shell masses and decay widths of the \PW and \PZ\ bosons provided by the Particle Data Group \cite{Tanabashi:2018oca},
\begin{equation}
  \begin{aligned}
    \MW^\OS&=80.379\,\GeV, & \GW^\OS&=2.085\,\GeV,\\
    \MZ^\OS&=91.1876\,\GeV, & \GZ^\OS&=2.4952\,\GeV,
  \end{aligned}
\end{equation}
to determine the pole masses and widths using the well-known formulae,
\begin{equation}
  M_V = \frac{1}{\sqrt{1+\left(\nicefrac{\Gamma_V^\OS}{M_V^\OS}\right)^2}}M_V^\OS,\quad
  \Gamma_V = \frac{1}{\sqrt{1+\left(\nicefrac{\Gamma_V^\OS}{M_V^\OS}\right)^2}}\Gamma_V^\OS,\quad
  V=\PW,\PZ.
\end{equation}
Furthermore, we employ the following mass and width parameters for the Higgs boson \cite{deFlorian:2016spz},
\begin{equation}
  \begin{aligned}
    \MH&=125\,\GeV,&\GH&=4.088\,\MeV,
  \end{aligned}
\end{equation}
and the top quark \cite{Tanabashi:2018oca},
\begin{equation}
  \Mt=173\,\GeV.
\end{equation}
The top quark only appears in closed fermion loops at NLO EW as we neglect mixing with the third generation quarks. Therefore, we can safely neglect the width of the top quark and assume it to be a stable particle, i.e.\ $\Gamma_\Pqt=0$.
All other fermions are assumed to be massless.
This, in particular, means that all leptons, including the \Pt lepton, are considered massless.
As we neglect the mixing involving quarks of the third generation, the CKM matrix factorizes from all matrix elements and can therefore be absorbed into the parton luminosities.
Furthermore, in this case, the SM is a CP-conserving theory, and the mixing among the first two generations is described by Cabibbo mixing with the Cabibbo angle
\begin{equation}
  \thetac = 0.22731.
\end{equation}
We apply the \Gmu-scheme \cite{Dittmaier:2001ay} where the electromagnetic coupling constant $\alpha$ is derived from the Fermi constant \cite{Tanabashi:2018oca},
\begin{equation}
  \Gmu=1.1663787\times10^{-5}\,\GeV^{-2},
\end{equation}
and given by
\begin{equation}
  \alpha = \alphagmu = \frac{\sqrt{2}}{\pi}\Gmu\MW\left(1-\frac{\MW^2}{\MZ^2}\right).
\end{equation}
In the \Gmu-scheme, the fine-structure constant at zero momentum transfer, $\alpha(0)$, is effectively evolved to the electroweak scale, thereby resumming large fermion-mass logarithms.
Additionally, leading universal two-loop correction to the $\rho$-parameter are absorbed into LO.
To prevent double counting, the charge renormalization constant $\delta Z_e$ defined via the Thomson limit has to be modified to
\begin{equation}
  \delta Z_e^{\Gmu} = \delta Z_e-\frac{1}{2}\Delta r,
\end{equation}
where $\Delta r$ comprises the quantum corrections to the muon decay \cite{Sirlin:1980nh,Denner:1991kt}.

To evaluate the $\Pp\Pp$ cross section, we chose a dynamical renormalization \mur and factorization scale \muf,
\begin{equation}
  \mur^2 = \muf^2 = \Big(3\MW\Big)^2+\Bigg(\sum_{i\in S}\mathbf{p}_{\rT,i}\Bigg)^{\!2},
\end{equation}
where the sum over $i$ runs over the vectorial transverse momenta $\mathbf{p}_{\rT,i}$ of all color-neutral particles $S$.
This scale choice is equal to the threshold energy for the production of three massive \PW bosons if there are no color-charged particles in the final state.
We use \propername{LHAPDF 6} \cite{Buckley:2014ana} to evaluate the parton distribution functions (PDFs).
In detail, we calculate the pure LO cross section \slo with the \propername{NNPDF 3.1 LO} \cite{Ball:2017nwa} and all NLO contributions, including the LO contribution $\slo_1$  to the NLO cross section, with the \propername{NNPDF 3.1 QCD+QED NLO} PDF set \cite{Bertone:2017bme}. The latter PDF set includes the photon PDF based on the \propername{LUXqed} approach \cite{Manohar:2016nzj,Manohar:2017eqh}.
Throughout all calculations we use the \alphas evolution given by the PDF set with
\begin{equation}
  \alphas(\MZ)=0.118.
\end{equation}

Using the definition of the $R$ distance of particles $i$ and $j$,
\begin{equation}
  \Delta R(i,j)=\sqrt{\Delta\eta_{ij}^2+\Delta\phi_{ij}^2},
\end{equation}
with differences of pseudorapidities $\Delta\eta_{ij}$ and azimuthal angles $\Delta\phi_{ij}$, we define a fiducial phase-space region inspired by the ATLAS and CMS experiments by demanding that the transverse momentum $p_\rT(\Pl)$ of each lepton \Pl and the $R$ distance of all pairs of leptons $\Pl_i$, $\Pl_j$ fulfill
\begin{equation}
  p_\rT(\Pl)>20\,\GeV,\qquad\Delta R(\Pl_i,\Pl_j)>0.1.
\end{equation}
Furthermore, the leading lepton $\Pl_1$, i.e.\ the one with the largest $p_\rT$, has to satisfy the condition
\begin{equation}
  p_\rT(\Pl_1)>27\,\GeV,
  \label{eq:parameters_l1cut}
\end{equation}
which is motivated by the lepton triggers of the LHC experiments.
Additionally, due to detector coverage, we demand
\begin{equation}
  \abs{\eta(\Pl)}<2.5.
\end{equation}
We recombine real-emitted photons with the nearest lepton \Pl, i.e.\ the lepton with the smallest $R$ distance to the photon, if
\begin{equation}
  \Delta R(\gamma,\Pl)<0.1
\end{equation}
to define collinear-safe observables.
This corresponds to the notion of dressed leptons used by the ATLAS and CMS experiments.

For 8 \TeV results presented in Tab.~\ref{tab:results_numresults8}, the leading-lepton $p_\rT$ requirement is dropped.
\subsection{Integrated cross sections}
\label{sec:results_offshellint}
We define the relative NLO corrections
\begin{align}
  \deltaqqew &\equiv \frac{\Delta\sigma^{\NLO\,\EW}_{\Pq\Paq^\prime}}{\sigma^\LO_1}, &
  \deltaqaew &\equiv \frac{\Delta\sigma^{\NLO\,\EW}_{\Pq\Pphot}}{\sigma^\LO}, &
  \deltaqcd &\equiv \frac{\sigma^{\LO}_1+\Delta\sigma^{\NLO\,\QCD}}{\sigma^\LO}-1,
  \label{eq:delta}
\end{align}
where the subscripts $\Pq\Paq^\prime$, $\Pq\Pphot$ indicate the partonic channels, i.e.\ quark--antiquark induced and quark--photon induced, respectively.
Normalizing the EW corrections $\Delta\sigma^{\NLO\,\EW}_{\Pq\Paq^\prime}$ to the LO cross section $\sigma^\LO_1$ evaluated with NLO PDFs, the relative EW correction $\deltaqqew$ is very insensitive to the PDF choice and depends on the factorization scale only very weakly.
The term $1+\deltaqcd$ corresponds to the usual definition of the QCD $K$-factor up to small QED corrections stemming from the PDFs owing to the normalization of the NLO QCD cross section to the LO cross section $\sigma^\LO$ evaluated with LO PDFs.
Combining the corrections multiplicatively, we define the full NLO relative correction $\delta^\NLO$,
\begin{equation}
  1+\delta^\NLO\equiv\big(1+\deltaqqew\big)\big(1+\deltaqcd\big)+\deltaqaew,
\end{equation}
so that
\begin{equation}
  \snlo = \bigl(1+\delta^\NLO\bigr) \times \slo.
\end{equation}
Analogous definitions will be used for differential cross sections $\mathrm{d}\sigma$ in the next subsection.

\begin{table}
  \centering
  \caption{LO and NLO cross sections, \slo and \snlo, as well as relative NLO corrections \deltaqqew, \deltaqaew, and \deltaqcd at different CM energies $\sqrt{s}$ of a proton--proton collision.
    Monte Carlo integration errors are indicated in parentheses.}
  \label{tab:results_numresults1314}
  \begin{subtable}{\linewidth}
    \centering
    \caption{$\Pp\Pp\rightarrow\Pem\Pane\Pmp\Pnm\Ptp\Pnt+X$}
    \begin{tabular}{d{2}d{1.9}d{1.7}d{+1.2}d{1.2}d{2.4}}
      \toprule
      \ccol{$\sqrt{s}$ {\small[TeV]}} & \ccol{$\sigma^\LO$ {\small[fb]}} & \ccol{$\sigma^\NLO$ {\small[fb]}} & \ccol{\deltaqqew\ {\small[\%]}}  & \ccol{\deltaqaew\ {\small[\%]}} & \ccol{\deltaqcd\ {\small[\%]}}\\
      \midrule
      13 & 0.194990(19) & 0.2626(10) & -7.7(4) & 7.22 & 38.02(4)\\
      14 & 0.20982(2) & 0.2872(12) & -7.8(4) & 7.78 & 40.04(4)\\
      \bottomrule
    \end{tabular}
  \end{subtable}\\
  \vspace*{1em}
  \begin{subtable}{\linewidth}
    \centering
    \caption{$\Pp\Pp\rightarrow\Pep\Pne\Pmm\Panm\Ptm\Pant+X$}
    \begin{tabular}{d{3}d{1.9}d{1.6}d{+1.2}d{1.2}d{2.4}}
      \toprule
      \ccol{$\sqrt{s}$ {\small[TeV]}} & \ccol{$\sigma^\LO$ {\small[fb]}} & \ccol{$\sigma^\NLO$ {\small[fb]}} & \ccol{\deltaqqew\ {\small[\%]}}  & \ccol{\deltaqaew\ {\small[\%]}} & \ccol{\deltaqcd\ {\small[\%]}}\\
      \midrule
      13 & 0.118411(12) & 0.1597(6) & -7.0(3) & 7.26 & 37.17(4)\\
      14 & 0.129986(13) & 0.1779(7) & -7.2(4) & 7.73 & 39.15(4)\\
      \bottomrule
    \end{tabular}
  \end{subtable}
\end{table}
\begin{table}
  \centering
  \caption{LO and NLO cross section, \slo and \snlo, and relative corrections \deltaqqew, \deltaqaew, and \deltaqcd for proton--proton collision at a CM energy of $\sqrt{s}=8\,\TeV$ without the additional phase-space cut \eqref{eq:parameters_l1cut} on the transverse momentum of the leading lepton. Monte Carlo integration errors are given in parentheses.}
  \label{tab:results_numresults8}
  \begin{tabular}{cd{1.9}d{1.7}d{1.2}d{1.2}d{2.4}}
    \toprule
    process & \ccol{$\sigma^\LO$ {\small[fb]}} & \ccol{$\sigma^\NLO$ {\small[fb]}} & \ccol{\deltaqqew\ {\small[\%]}}  & \ccol{\deltaqaew\ {\small[\%]}} & \ccol{\deltaqcd\ {\small[\%]}}\\
    \midrule
    $\Pp\Pp\rightarrow\Pem\Pane\Pmp\Pnm\Ptp\Pnt+X$ & 0.114614(11) & 0.1405(5) & -6.8(3) & 4.27 & 26.91(3)\\
    $\Pp\Pp\rightarrow\Pep\Pne\Pmm\Panm\Ptm\Pant+X$ & 0.060673(6) & 0.0744(2) & -6.2(3) & 4.56 & 25.96(4)\\
    \bottomrule
  \end{tabular}
\end{table}
We present LO and NLO cross sections, as well as the NLO corrections for the two charge-conjugated processes of $\PW\PW\PW$ production, $\Pp\Pp\rightarrow\Pem\Pane\Pmp\Pnm\Ptp\Pnt+X$ and $\Pp\Pp\rightarrow\Pep\Pne\Pmm\Panm\Ptm\Pant+X$, for the current and the planned LHC CM energies of 13 and 14\,\TeV in Tab.\ \ref{tab:results_numresults1314}.
Due to the high power in the EW coupling constant $\alpha$ the cross sections are fairly small.
The QCD corrections dominate the NLO corrections and amount to $\sim$\,38--40\,\% at the current and upcoming CM energies of the LHC of 13/14\,\TeV.
Similarly to Ref.\ \cite{Dittmaier:2017bnh}, where on-shell $\PW\PW\PW$ production was analyzed, we observe a large cancellation between the quark--photon and the quark--antiquark-induced NLO EW corrections.
Within the chosen parameter and event-selection setup, they are of the same size, but have opposite sign.
This cancellation is not systematic, i.e.\ the two types of corrections are uncorrelated.
The quark--photon-induced channels are highly sensitive to a potential jet veto, as was already shown in Ref.~\cite{Dittmaier:2017bnh}.
This is similarly valid for the QCD corrections.
The cross sections of $\PWm\PWp\PWp$ production are approximately 1.6 times larger than the ones of its charge-conjugated process, but the relative corrections are nearly identical.
As the SM within our parameter set is a CP-conserving theory, the difference between the cross sections for $\PWm\PWp\PWp$ and $\PWp\PWm\PWm$ purely arises due to the difference in the PDFs of quarks and antiquarks.
Additional results for a CM energy of 8\,\TeV are presented in Tab.\ \ref{tab:results_numresults8}, where we have dropped the extra requirement \eqref{eq:parameters_l1cut} on the transverse momentum of the leading lepton.
Due to the lower CM energy of the collider, and therefore smaller average partonic energy, the cross sections are significantly smaller than the ones for the higher collider CM energies.
Both relative QCD and EW corrections are somewhat smaller than for the scattering energies of 13/14\,\TeV.
For the EW $\Pq\Pphot$ contributions and the QCD corrections, which are dominated by the real corrections, this is due to the smaller phase-space for real-particle emission at $8\,\TeV$;
for the EW $\Pq\Paq^\prime$ contributions the corrections for $13/14\,\TeV$ are more negative because of the deeper reach into regions of high partonic scattering energies where the EW corrections grow large and negative due to EW Sudakov logarithms (see differential cross sections in Sec.~\ref{sec:results_offshelldiff}).

\begin{table}
  \caption{Scale uncertainty for the process $\Pp\Pp\rightarrow\Pem\Pane\Pmp\Pnm\Ptp\Pnt+X$ at a CM energy of $13\,\TeV$ obtained by varying the renormalization and the factorization scale together up- and down by a factor of two.}
  \label{tab:results_numresultsscale}
  \centering
  \begin{tabular}{cc}
    \toprule
    \ccol{$\sigma^\LO$ {\small[fb]}} & \ccol{$\sigma^\NLO$ {\small[fb]}} \\
    \midrule
    0.1950$^{+1.0\%}_{-1.5\%}$ & 0.2626$^{+2.4\%}_{-2.8\%}$\\
    \bottomrule
  \end{tabular}
\end{table}
In Tab.\ \ref{tab:results_numresultsscale} we present the scale uncertainties for the $\PWm\PWp\PWp$ production process $\Pp\Pp\rightarrow\Pem\Pane\Pmp\Pnm\Ptp\Pnt+X$ obtained by varying the factorization and the renormalization scales, $\muf$ and $\mur$, together up and down by a factor of two.
At LO, the cross section of the given process is a pure EW quantity, i.e.\ no powers in \alphas are present in the transition matrix element, so that the scale uncertainty does not reflect the size of the theoretical uncertainties due to missing higher-order corrections.
Only at NLO QCD, renormalization scale dependent contributions arise, which do not decrease the small scale uncertainties of the LO cross sections.
This observation is in agreement with the results for the on-shell calculation of Ref.~\cite{Dittmaier:2017bnh}.

\subsection{Differential cross sections}
\label{sec:results_offshelldiff}
In the following, we present differential distributions for the process $\Pp\Pp\rightarrow\Pem\Pane\Pmp\Pnm\Ptp\Pnt+X$ at a CM energy of $\sqrt{s}=13\,\TeV$.
To this end, we define the missing momentum $p_\miss$ as the sum of the momenta of the neutrinos and antineutrinos, and the momenta of unidentified particles,
\begin{equation}
  p_\miss=\sum_ip_{\nu_i}+\sum\limits_{\substack{k\neq \nu\\\text{$k$ unid.}}}p_k,
\end{equation}
where a particle $k$ after recombination is ``unidentified'' if it does not pass the identification criteria
\begin{equation}
  p_T(k) > 20\,\GeV,\qquad \lvert\eta(k)\rvert < 5.
\end{equation}
The missing transverse energy $E_{\rT,\miss}$ is defined as the absolute value of the transverse part of the missing momentum $p_\miss$.

\begin{figure}
  \centering
  \includegraphics{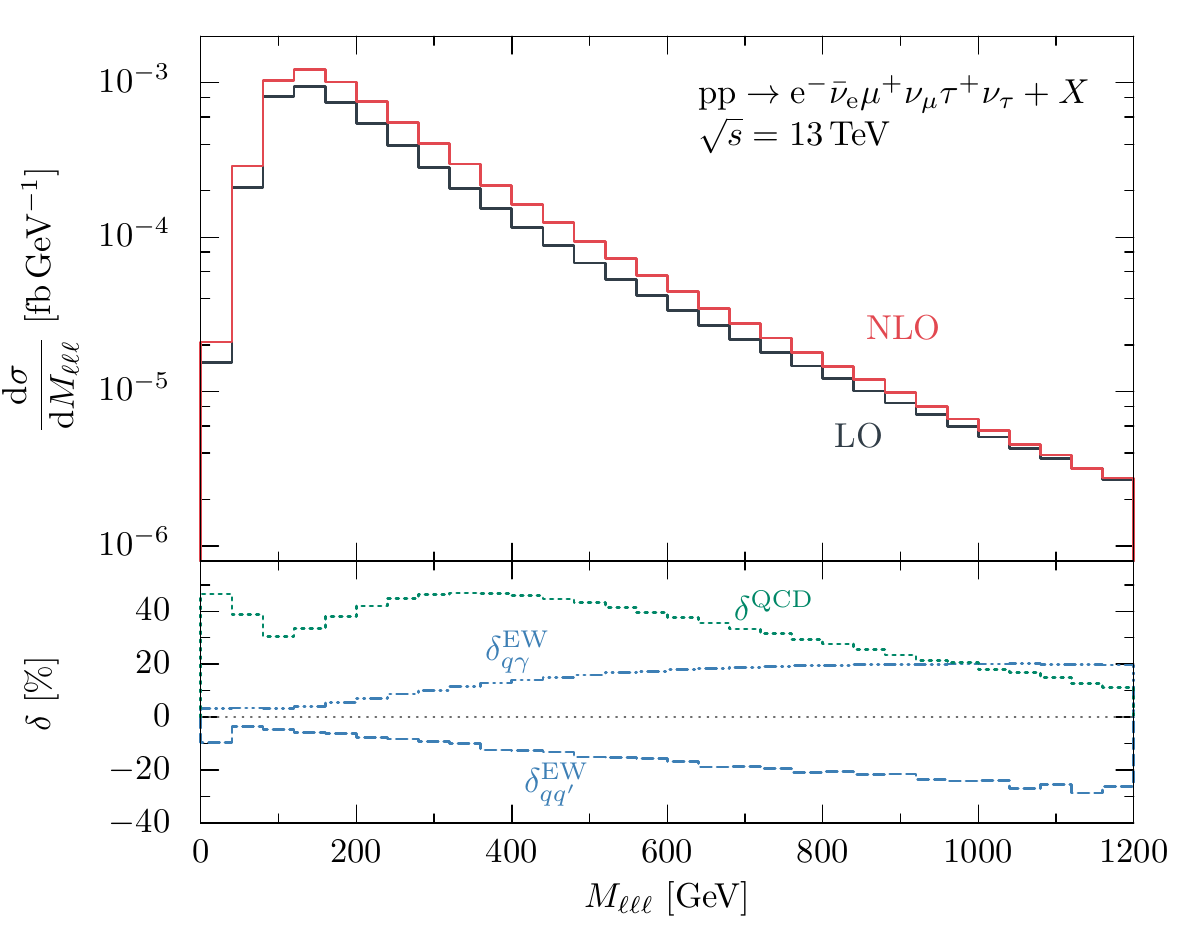}
  \caption{Differential LO and NLO cross section, and relative NLO corrections in the invariant mass $M_{\ell\ell\ell}$ of the three-lepton system. The upper part visualizes the dependence of the LO and NLO cross sections of $M_{\Pl\Pl\Pl}$. The lower part shows the relative NLO corrections. }
  \label{fig:results_mlll}
\end{figure}
Figure \ref{fig:results_mlll} shows differential distributions of the LO and NLO corrections as well as the NLO corrections in the invariant mass $M_{\ell\ell\ell}$ of the three-lepton system.
While the quark--photon-induced EW corrections increase with growing invariant mass, the quark--antiquark-induced correction decrease down to $\sim-30\,\%$ in the TeV range.
This is due to the strong impact of EW high-energy logarithms.

\begin{figure}
  \centering
  \includegraphics{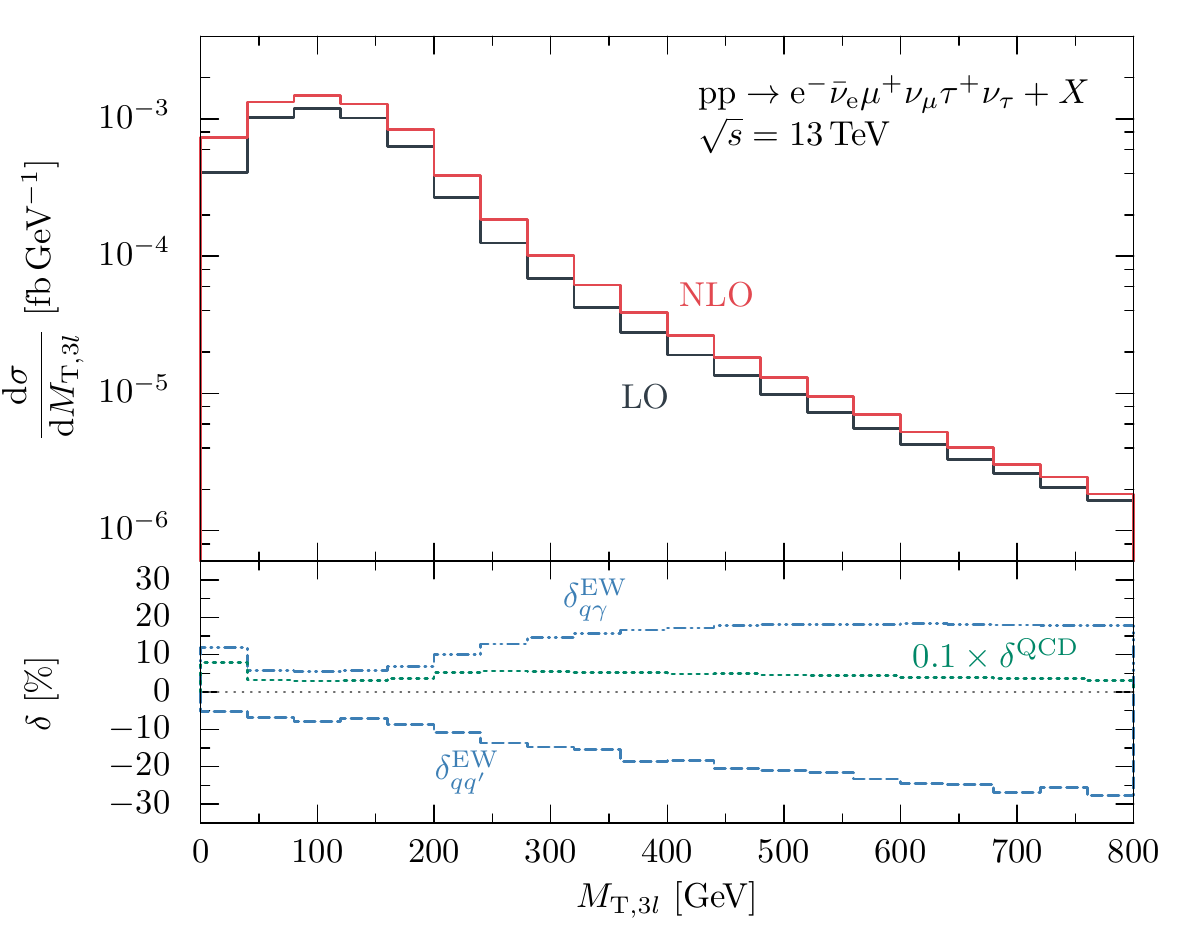}
  \caption{Differential distribution in the transverse mass $M_{\rT,3\ell}$ of the three-lepton system. The NLO QCD correction $\delta^\QCD$ is scaled down by a factor of 10 for better readability.}
  \label{fig:results_mt3l}
  \vspace*{\floatsep}
  \includegraphics{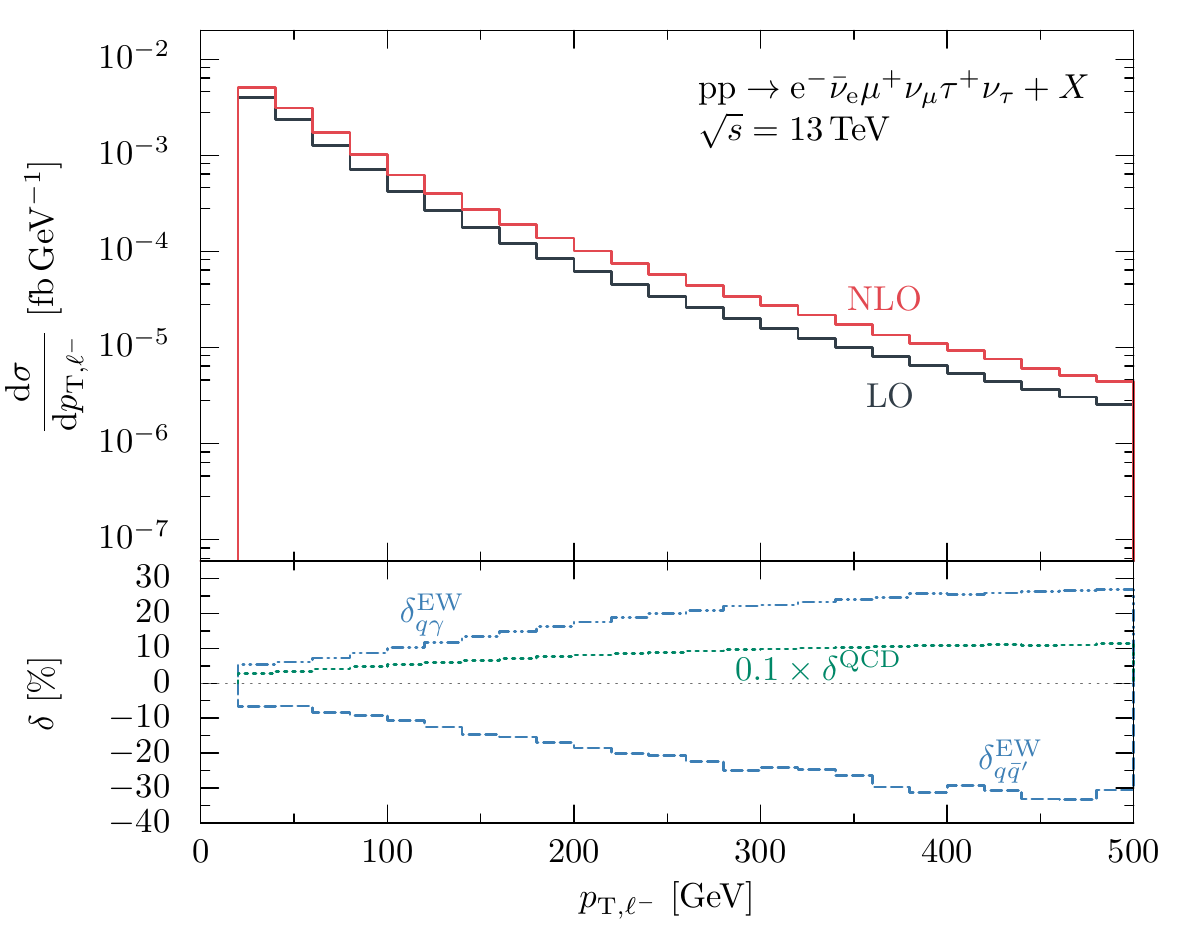}
  \caption{Differential distribution in the transverse momentum $p_{\rT,\Plm}$ of the negatively charged lepton in $\Pem\Pane\Pmp\Pnm\Ptp\Pnt+X$ production.}
  \label{fig:results_ptl-}
\end{figure}

In Fig.~\ref{fig:results_mt3l} we present differential distributions in the transverse mass of the three-lepton system defined via
\begin{equation}
  M_{\rT,3\ell}\equiv \sqrt{2 p_{\rT}(3\ell) E_{\rT,\miss} \Bigl[1 -\cos\bigl(\Delta\phi_{p_{3\ell}p_\miss}\bigr)\Bigr]},
\end{equation}
where $p_{3\ell}$ is the momentum of the three-lepton system, $p_{\rT}(3\ell)$ the transverse part of $p_{3\ell}$, and $\Delta\phi_{p_{3\ell}p_\miss}$ the azimuthal angle difference of the missing momentum and the three-lepton system in the plane transverse to the beams.
We observe a strong influence of negative high-energy logarithms similarly to the three-lepton invariant mass at large values of $M_{\rT,3\ell}$.

We show transverse momentum distributions of the negatively charged lepton in Fig.~\ref{fig:results_ptl-}. The effect of negative high-energy logarithms becomes apparent in regions of large missing transverse momentum, similarly as was observed for the $M_{\Pl\Pl\Pl}$ and $M_{\rT,3\ell}$ distributions. The large impact of QCD corrections is due to recoil effect from hard jet emission where the whole $\PW\PW\PW$ system receives a strong boost transverse to the beams.
This effect is also well known from \PW 
and $\PW\PW$ 
production processes and could be reduced by a jet veto.

\begin{figure}
  \centering
  \includegraphics{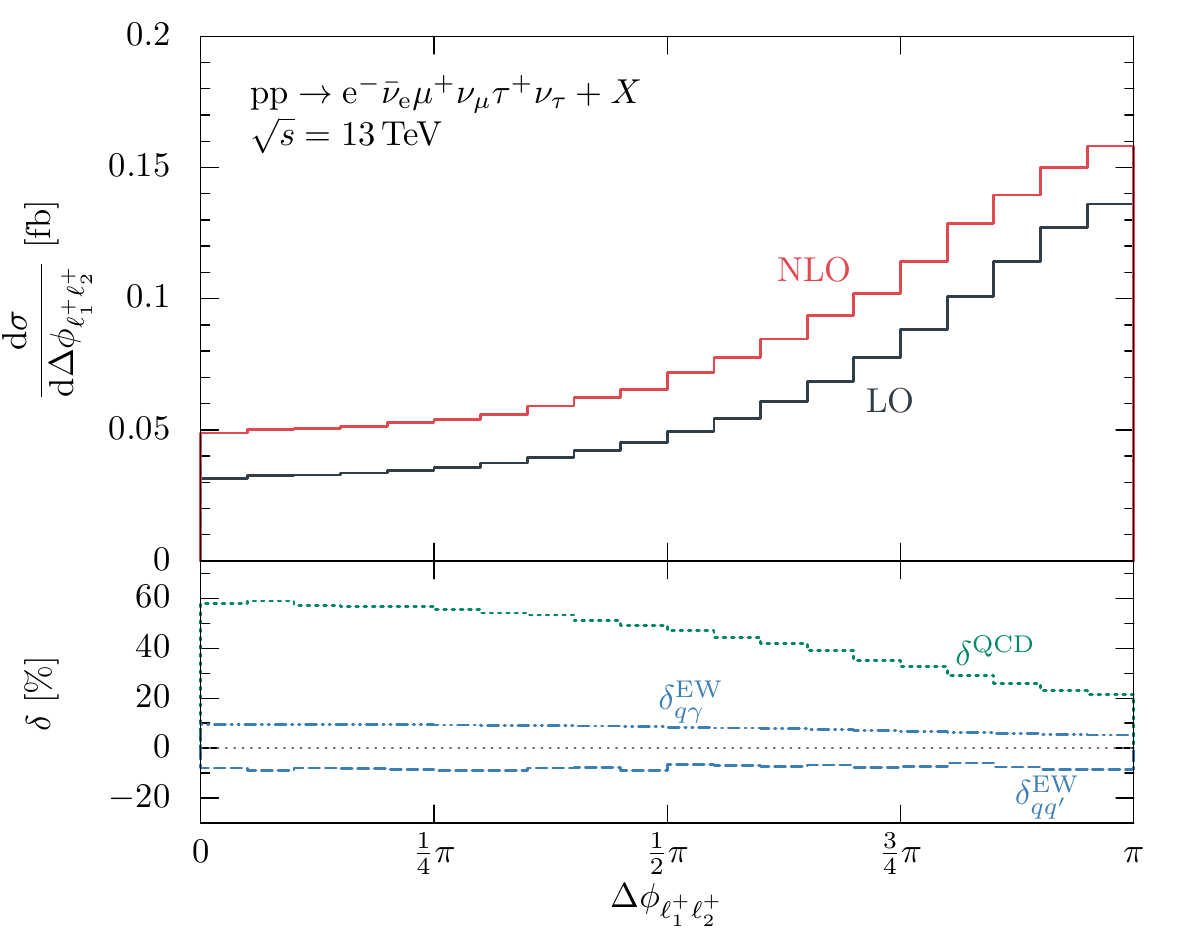}
  \caption{Differential LO and NLO cross section and relative NLO corrections in the difference in the azimuthal angle of the two positively charged leptons, $\Delta\phi_{\ell^+_1\ell^+_2}$.}
  \label{fig:results_deltaphi}
  \vspace*{\floatsep}
  \centering
  \includegraphics{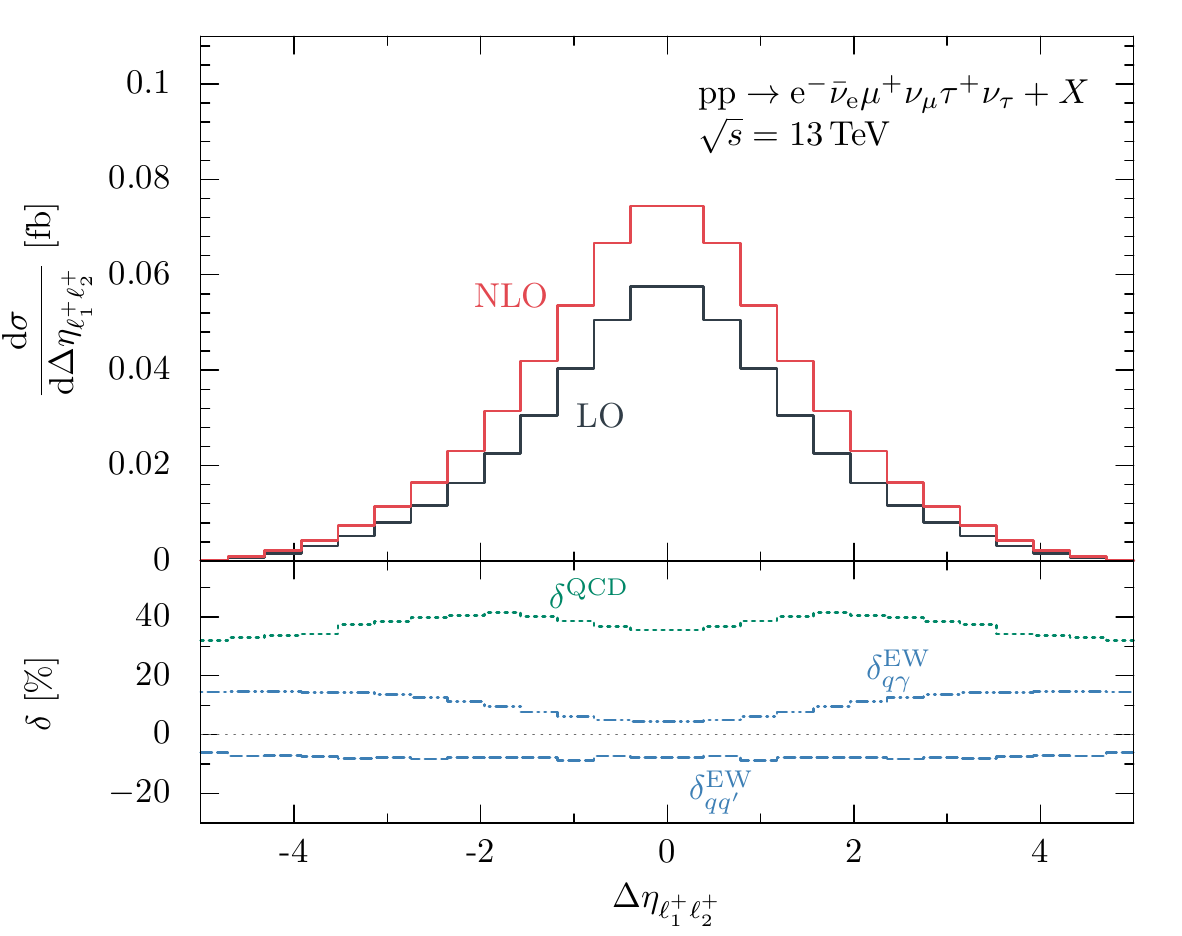}
  \caption{Differential distribution in the difference of the pseudorapidities of the two positively charged leptons, $\Delta\eta_{\ell^+_1\ell^+_2}$.}
  \label{fig:results_deltaeta}
\end{figure}

Figures \ref{fig:results_deltaphi} and \ref{fig:results_deltaeta} depict important angular and pseudorapidity distributions which might be crucial in the search for anomalous gauge couplings.
The EW corrections do not distort the shape of the differential distribution in the difference of the azimuthal angle of the two positively charged leptons, $\Delta\phi_{\ell^+_1\ell^+_2}$, and are fairly universal.
However, the QCD corrections amount to nearly 60\% in the low-$\Delta\phi$ region, while in the high-$\Delta\phi$ region the correction is only about 20\%.
We attribute this enhanced impact of QCD corrections again to the recoil of the leptons in the case of hard jet emission, which reduces the angles between the leptons by a strong boost transverse to the beams.
We observe that the quark--antiquark-induced EW corrections $\deltaqqew$ are quite independent of the difference in the pseudorapidities of the two positively charged leptons, $\Delta\eta_{\ell^+_1\ell^+_2}$.
The other NLO corrections somewhat distort the shape of the distribution.

\subsection{Comparison of results from the triple-pole approximation with results of the full off-shell calculation}
\label{sec:results_tpa-vs-full}
Here, we compare cross sections of $\Pem\Pane\Pmp\Pnm\Ptp\Pnt$ production evaluated within the TPA with results of the full off-shell calculation.
For a meaningful comparison, we further restrict the phase space by excluding the Higgs resonance in the off-shell evaluated contributions upon applying cuts on the invariant masses of $\PWm\PWp$ pairs,
\begin{equation}
  \abs{M_{\ell_i^+\Pn_i\ell_j^-\Pan_j}-\MH} > 1\,\GeV,
  \label{eq:parameters_hcut1}
\end{equation}
where $\ell_{i/j}^\pm$ is any lepton with charge $\pm e$ and $\Pn_i$, $\Pan_j$ the corresponding (anti)neutrino determined from ``Monte Carlo truth''.
If the external photon of the real calculation has not been recombined we further require
\begin{equation}
  \abs{M_{\ell_i^+\Pn_i\ell_j^-\Pan_j\gamma}-\MH} > 1\,\GeV.
  \label{eq:parameters_hcut2}
\end{equation}
These cuts cannot be realized in experimental analyses, which is acceptable here, because we are mainly interested in the comparison between full off-shell calculation and the TPA.
Since the exclusion of the Higgs resonance is rather minimalistic, the following results on the quality of the TPA certainly remain valid if a more realistic isolation of the Higgs-strahlung process is adopted.
Additionally, the results presented in this section could be combined with the known NLO cross sections for $\PW\PH$ production \cite{Denner:2011id,deFlorian:2016spz,Granata:2017iod} and $\PH\to\PW\PW$ decays \cite{Bredenstein:2006rh} from the literature to obtain good approximations for the cross sections of $\Pem\Pane\Pmp\Pnm\Ptp\Pant$ production.

We define the relative difference $\Delta_\TPA$ between the full off-shell calculation and the TPA by
\begin{equation}
  \Delta_\TPA\equiv\frac{\sigma^\NLO_\TPA-\sigma^\NLO_\full}{\sigma^\NLO_\full},
\end{equation}
where $\sigma^\NLO_\TPA$ is the NLO cross section evaluated in the TPA and $\sigma^\NLO_\full$ the NLO cross section evaluated fully off shell.
We recall that the LO parts of both NLO cross sections are based on full $2 \to 6$ off-shell matrix elements and are thus identical.
The TPA can either be applied to the EW correction, to the QCD correction, or simultaneously to both types of corrections.
Therefore, we additionally distinguish $\Delta_{\TPA,\EW}$, $\Delta_{\TPA,\QCD}$, and $\Delta_{\TPA}$, respectively.
We further define
\begin{equation}
  \Delta_\TPA^\LO\equiv\frac{\sigma^\LO_\TPA-\sigma^\LO_1}{\sigma^\LO_1},
\end{equation}
where $\sigma^\LO_\TPA$ is the LO cross section evaluated in the TPA with NLO PDFs.
Note that $\sigma^\LO_\TPA$ is a pure auxiliary quantity only needed to calculate $\Delta_\TPA^\LO$, which will be relevant in the discussion of the TPA accuracy.

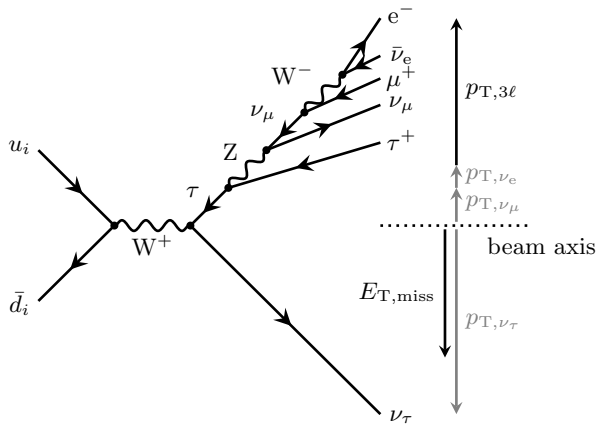
\begin{figure}
  \centering
  \begin{tikzpicture}
  [baseline={([yshift=-0.95ex]current bounding box.center)},
  line width=1pt,
  phot/.style={decorate,decoration={snake, segment length=#1, amplitude=2pt}},
  phot/.default=9pt,
  ferm/.style={postaction={decorate},decoration={markings, mark=at position #1 with {\stealtharrow}}},
  ferm/.default=0.5,
  higgs/.style=dashed,
  scale=0.5]
  \footnotesize
  \coordinate (i1) at (-1,2);
  \coordinate (i2) at (-1,-2);
  \coordinate (o1) at (8,-5);
  \coordinate (o2) at (8,2.2);
  \coordinate (o3) at (8,3.2);
  \coordinate (o4) at (8,3.9);
  \coordinate (o5) at (8,4.5);
  \coordinate (o6) at (8,5.5);
  \coordinate (v1) at (1,0);
  \coordinate (v2) at (3,0);
  \coordinate (v3) at (4,1);
  \coordinate (v4) at (5,2);
  \coordinate (v5) at (6,3);
  \coordinate (v6) at (7,4);
  \coordinate (beamc) at (10,0);
  \coordinate (pt3l) at ($(o2) + (o4) - (o3) + (o6) - (o5)$);
  \coordinate (ptnm) at ($(o3) - (o2)$);
  \coordinate (ptne) at ($(o5) - (o4)$);
  \coordinate (etmiss) at ($(o1)+(ptnm)+(ptne)$);
  \draw[ferm] (i1) -- (v1);
  \draw[ferm] (v1) -- (i2);
  \draw[phot] (v1) -- node[below]{\PWp} (v2);
  \draw[ferm] (o2) -- (v3);
  \draw[ferm] (v3) -- node[above left]{\Pt} (v2);
  \draw[ferm] (v2) -- (o1);
  \draw[phot] (v3) -- node[above left]{\PZ}(v4);
  \draw[ferm] (o4) -- (v5);   
  \draw[ferm] (v5) -- node[above left]{\Pnm}(v4);
  \draw[ferm] (v4) -- (o3);
  \draw[phot] (v5) -- node[above left]{\PWm}(v6);
  \draw[ferm] (o5) -- (v6);
  \draw[ferm] (v6) -- (o6);   
  \draw[fill=black] (v1) circle (2pt);
  \draw[fill=black] (v2) circle (2pt);
  \draw[fill=black] (v3) circle (2pt);
  \draw[fill=black] (v4) circle (2pt);
  \draw[fill=black] (v5) circle (2pt);
  \draw[fill=black] (v6) circle (2pt);
  \node[xshift=-7pt, yshift=1pt] at (i1) {$u_i$};
  \node[xshift=-7pt, yshift=-1pt] at (i2) {$\bar{d}_i$};
  \node[xshift=8pt, yshift=-1pt] at (o1) {\Pnt};
  \node[xshift=8pt, yshift=1pt] at (o2) {\Ptp};
  \node[xshift=8pt] at (o3) {\Pnm};
  \node[xshift=8pt] at (o4) {\Pmp};
  \node[xshift=8pt] at (o5) {\Pane};
  \node[xshift=8pt, yshift=2pt] at (o6) {\Pem};
  \draw[dotted] (beamc)+(-2,0) -- node[below right=1pt and 8pt]{beam axis}+(2,0);
  \draw[-stealth, gray] (beamc)+(0,-0.1) -- node[right]{$p_{\rT,\Pnt}$}(beamc |- o1);
  \draw[-stealth, gray] (beamc)+(0,0.1) -- node[right]{$p_{\rT,\Pnm}$}(beamc |- ptnm);
  \draw[-stealth, gray] (beamc |- ptnm) -- node[right]{$p_{\rT,\Pne}$} +(0,0 |- ptne);
  \draw[-stealth] ($(beamc |- ptnm)+(0,0 |- ptne)$) -- node[right]{$p_{\rT,3\Pl}$}+(0,0 |- pt3l);
  \draw[-stealth] (beamc)++(-0.3,-0.1) -- node[left]{$E_{\rT,\miss}$}+(0,0 |- etmiss);
\end{tikzpicture}
  \caption{Illustration of the diagrammatic structures relevant for the difference in the TPA and the full off-shell calculation for high missing transverse energies \etmiss.}
  \label{fig:tpavsfull_etmiss_illustration}
\end{figure}

We can estimate the size of $\Delta_\TPA$ by investigating the different contributions that have an impact on $\Delta_\TPA$ as discussed in Ref.~\cite{Biedermann:2016guo} for \PW-pair production.
In regions where the contributions with three resonant \PW bosons dominate the cross section, $\Delta_\TPA$ can naively be estimated to $\frac{\alpha}{\pi}\frac{\GW}{\MW}c_\EW\lesssim 0.5\,\%$ for the EW contribution and to $\frac{\alphas}{\pi}\frac{\GW}{\MW}c_\QCD\lesssim 0.5\,\%$ for the QCD contribution, where $c_{\EW/\QCD}$ are enhancement factors, resulting e.g.\ from double and single logs, estimated very conservatively.
This can be motivated by the fact that in the TPA we only take the leading term in the expansion of the NLO corrections about the three resonant \PW propagators, neglecting off-shell terms that are typically suppressed by a factor of $\nicefrac{\GW}{\MW}$ resulting in terms of the mentioned size.
In some regions of phase space, the cross sections become sensitive to off-shell contributions.
As already observed for \PW-pair production in Ref.~\cite{Biedermann:2016guo}, regions of large lepton-$p_\rT$ and $E_{\rT,\miss}$ are particularly prone to large off-shell effects owing to the enhanced contributions of diagrams as illustrated in Fig.~\ref{fig:tpavsfull_etmiss_illustration}, which are absent in the TPA and where single leptons recoil against all other produced leptons.
This enhancement can already be observed at LO, i.e.\ in the quantity $\Delta_\TPA^\LO$.
To estimate the size of $\Delta_\TPA$ in these regions, we propagate $\Delta_\TPA^\LO$ to NLO by multiplication with a suitable NLO correction.
As we apply the TPA solely to the virtual contributions, we multiply $\Delta_\TPA^\LO$ with measures $\Delta_\text{virt}^{\EW/\QCD}$ of the respective NLO EW and QCD corrections that stem from the virtual contributions,
\begin{align}
  \Delta_\text{virt}^\EW&\equiv\frac{\Delta\sigma^{\EW,\TPA}_{\text{virt.}+\bm{I}}}{\sigma^\LO_1}, & \Delta_\text{virt}^\QCD&\equiv\frac{\Delta\sigma^{\QCD,\TPA}_{\text{virt.}+\bm{I}}}{\sigma^\LO_1},
\end{align}
where the subscript indicates that we use the IR-finite contributions given on amplitude level by the virtual TPA one-loop amplitude plus the $\bm{I}$-operator/endpoint contributions from dipole subtraction.
Note that the given approach is well motivated as we solely apply the TPA to the virtual contributions and evaluate all other contributions fully off shell.
Using a more general approach by multiplication with e.g.\ the NLO corrections from the quark--antiquark-induced channels would strongly overestimate the uncertainty of the TPA in regions where the real emission contributions dominate over the virtual contributions. 
In total, the size of $\Delta_\TPA$ can be estimated to
\begin{equation}
  \abs{\Delta_\TPA}\sim\Delta_\TPA^\text{estimate}=\max\biggl\{
  \frac{\alpha}{\pi}\frac{\GW}{\MW}c_\EW,
  \frac{\alphas}{\pi}\frac{\GW}{\MW}c_\QCD,
  \Bigl\lvert\Delta_\TPA^\LO\Delta_\text{virt}^\EW\Bigr\rvert,
  \Bigl\lvert\Delta_\TPA^\LO\Delta_\text{virt}^\QCD\Bigr\rvert\biggr\}.
  \label{eq:tpavsfull_deltatpa_estimate}
\end{equation}
Note that for this estimate, we do not need to know the full off-shell NLO results.
It can be calculated from the TPA results and the additional auxiliary quantity $\sigma^\LO_\TPA$ only.

\begin{table}
  \centering
  \caption{Comparison of the NLO cross sections \snlo of the process $\Pp\Pp\rightarrow\Pem\Pane\Pmp\Pnm\Ptp\Pnt + X$ in the TPA with the full off-shell calculation and the relative differences $\Delta_\TPA$ for different CM energies  $\sqrt{s}$.
    Monte Carlo integration errors are given in parentheses.}
  \label{tab:tpavsfull_int_cm}
  \begin{tabular}{d{2}d{1.7}d{1.7}d{1.2}d{1.2}d{1.2}d{1.1}}
    \toprule
    \ccol{\multirow{2}{*}{$\sqrt{s}$ {\small[TeV]}}} & \multicolumn{2}{c}{$\sigma^\NLO$ {\small [fb]}} & \ccol{\multirow{2}{*}{$\Delta_\TPA$ {\small[\%]}}} & \ccol{\multirow{2}{*}{$\Delta_{\TPA,\EW}$ {\small[\%]}}} & \ccol{\multirow{2}{*}{$\Delta_{\TPA,\QCD}$ {\small[\%]}}} & \ccol{\multirow{2}{*}{$\Delta_{\TPA}^\text{estimate}$ {\small[\%]}}}\\
    \cline{2-3}
     & \ccol{TPA} & \ccol{full}\\
    \midrule
    13 & 0.14581(4) & 0.14572(4) & 0.06 & 0.47 & -0.40 & 0.8\\
    14 & 0.16143(4) & 0.16130(5) & 0.08 & 0.51 & -0.42 & 0.8\\
    \bottomrule
  \end{tabular}
\end{table}
\begin{table}
  \centering
  \caption{Comparison of the LO and the NLO cross sections, \slo and \snlo, and the relevant NLO corrections \deltaqqew and \deltaqcd for the TPA and the full off-shell calculation for the process $\Pp\Pp\rightarrow\Pem\Pane\Pmp\Pnm\Ptp\Pnt + X$ at a CM energy of $\sqrt{s}=13\,\TeV$.
  Monte Carlo integration errors are indicated in parentheses.}
  \label{tab:tpavsfull_int_detail}
  \begin{tabular}{cd{1.9}d{1.7}d{+2.4}d{2.4}}
    \toprule
    & \ccol{$\sigma^\LO$ {\small[fb]}} & \ccol{$\sigma^\NLO$ {\small[fb]}} & \ccol{\deltaqqew\ {\small[\%]}}  & \ccol{\deltaqcd\ {\small[\%]}}\\
    \midrule
    TPA  & 0.093436(5) & 0.14581(4) & -7.80(2) & 57.82(3)\\
    full & 0.093436(5) & 0.14572(4) & -8.26(2) & 58.50(3)\\
    \bottomrule
  \end{tabular}
\end{table}%
As can be seen in Tab.\ \ref{tab:tpavsfull_int_cm}, the integrated cross sections in the TPA are in very good agreement with the off-shell result for different CM energies.
To some extent, this agreement stems from a cancellation between the NLO QCD and NLO EW contributions:
The effect of applying the TPA to the EW corrections compensates the effect of applying the TPA to the QCD corrections (see Tab.\ \ref{tab:tpavsfull_int_detail}).
The relative difference of the off-shell calculation and applying the TPA only in the calculation of the NLO EW corrections amounts for $0.47\,\%$ at a CM energy of 13\,\TeV, while for the QCD corrections we obtain $0.40\,\%$.
This is in agreement with the naive error estimate of the TPA.
The estimate $\Delta_\TPA^\text{estimate}$ somewhat overestimates the difference between the full off-shell calculation and the calculation within the TPA because of a mediocre performance of the TPA at LO, $\Delta_\TPA^\LO\sim 3\,\%$, and sizable QCD corrections, $\Delta_\text{virt}^\QCD\sim28\,\%$.
Nevertheless, it captures the uncertainty of the TPA well.

\begin{figure}[htbp]
  \centering
  \includegraphics[scale=0.95]{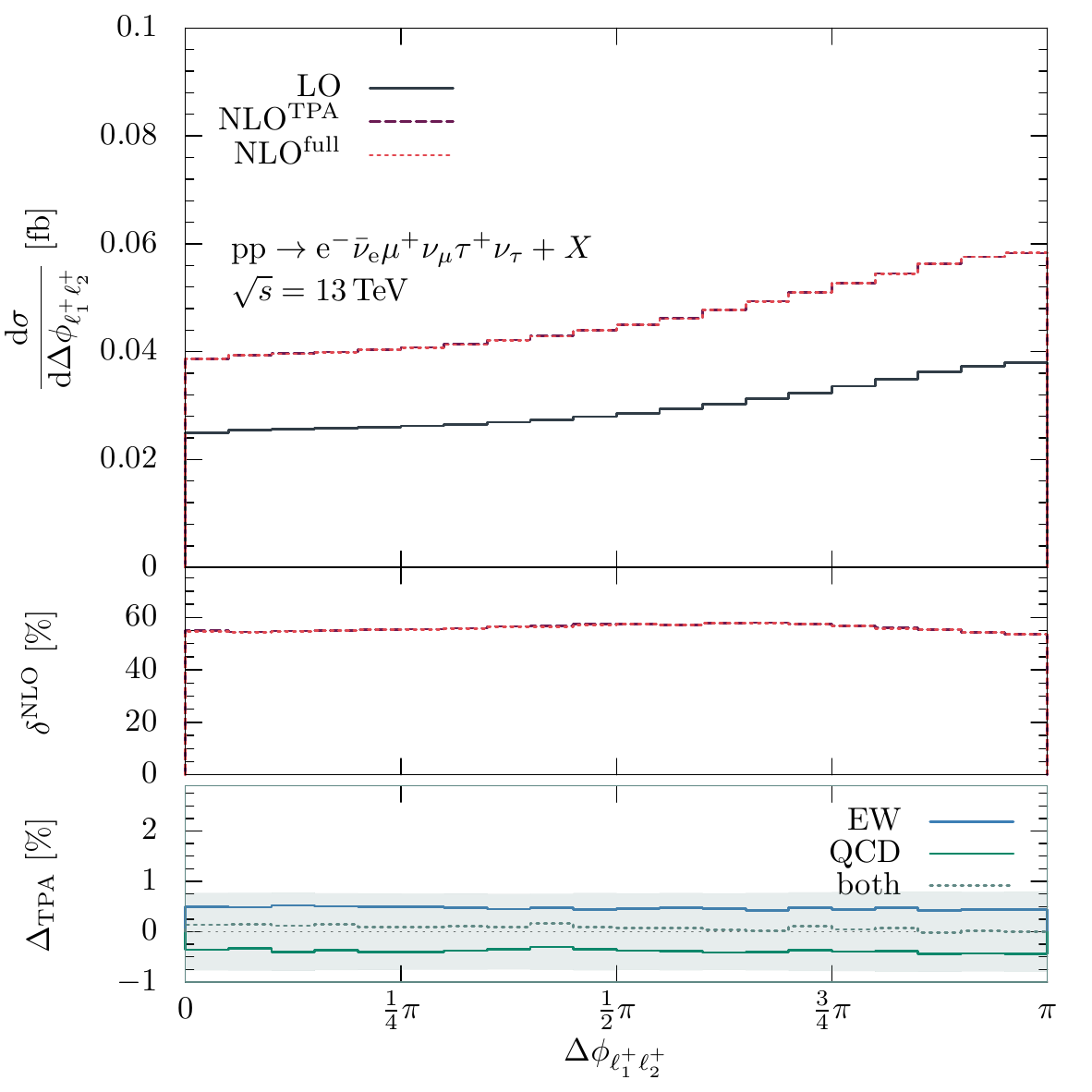}
  \caption{Differential cross sections and relative corrections over the azimuthal-angle difference $\Delta\phi_{\ell^+_1\ell^+_2}$ of the two positively-charged leptons.
    The upper part shows the differential LO cross sections and the NLO cross sections for the TPA and the full off-shell calculation.
    In the middle, the total NLO correction $\delta^\NLO$ for the TPA and the full off-shell calculation are depicted.
    The bottom panel shows the relative difference $\Delta_\TPA$ between the TPA and the full off-shell calculation if taking the TPA only in the NLO EW contributions, the NLO QCD contribution, or concurrently in both parts. The shaded gray area indicates the estimated size of $\Delta_\TPA$, $\Delta_\TPA^\text{estimate}$, following Eq.~\eqref{eq:tpavsfull_deltatpa_estimate}.}
  \label{fig:tpavsfull_deltaphi}
\end{figure}
\begin{figure}[p]
  \centering
  \includegraphics[trim=0 0.15cm 0 0.15cm,clip,scale=0.95]{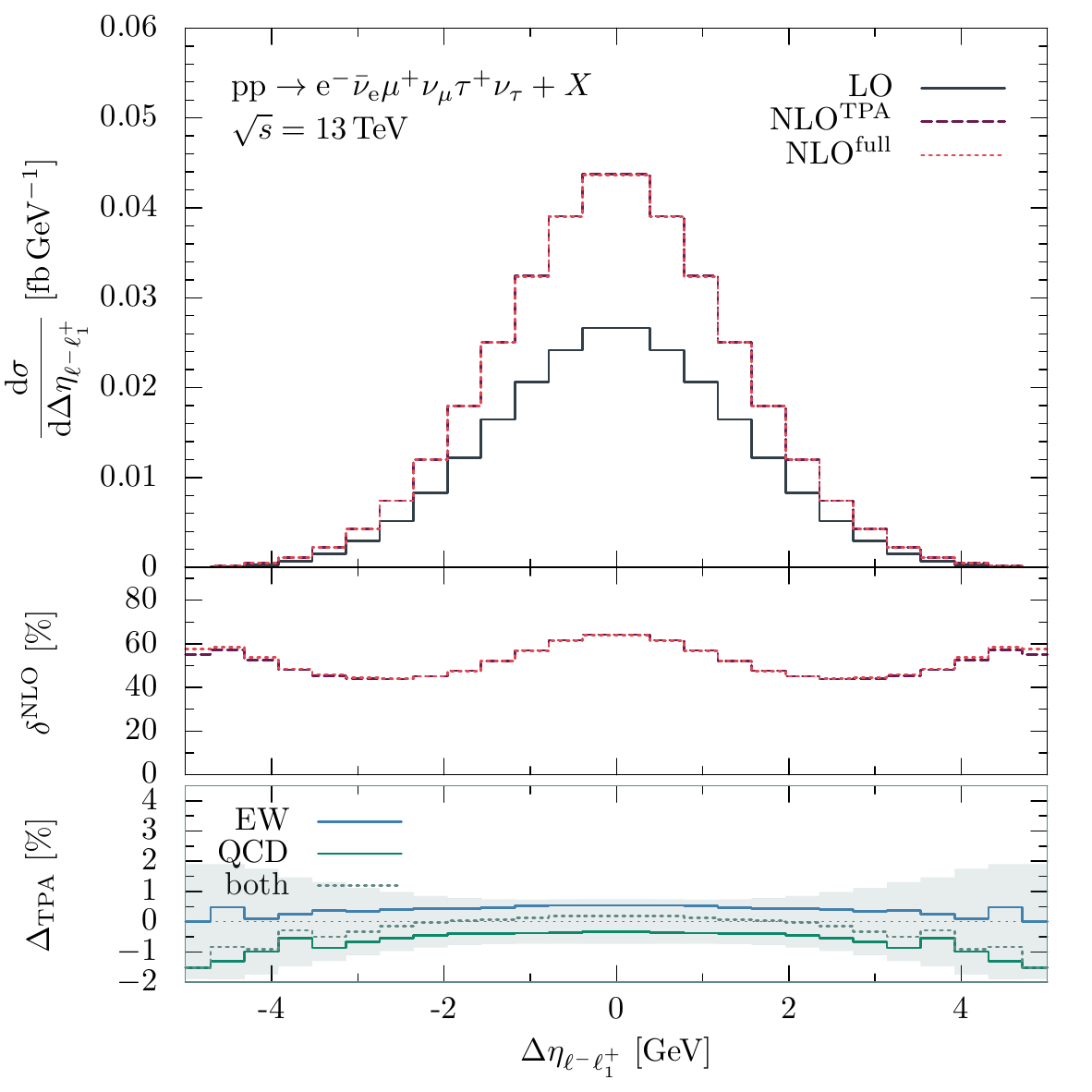}
  \caption{As in Fig.~\ref{fig:tpavsfull_deltaphi}, but for the differential distributions in the pseudorapidity difference $\Delta\eta_{\Plm\ell^+_1}$ of the negatively-charged lepton $\Plm$ and the leading positively-charged lepton $\ell^+_1$.}
  \label{fig:tpavsfull_deltaeta}
\end{figure}
\begin{figure}[p]
  \centering
  \includegraphics[trim=0 0.25cm 0 0.1cm,clip,scale=0.95]{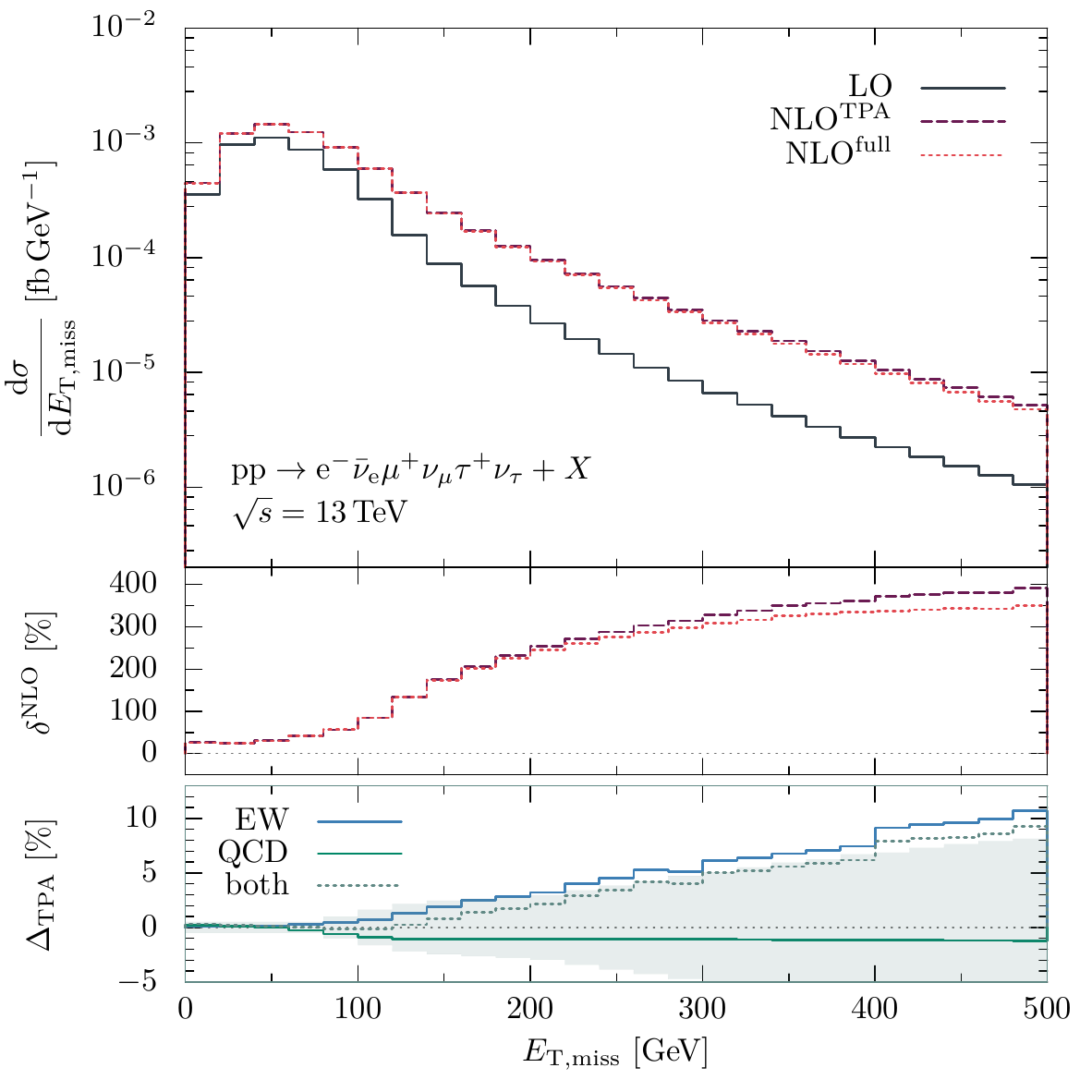}
  \caption{As in Fig.~\ref{fig:tpavsfull_deltaphi}, but for the distribution in the missing transverse momentum \etmiss.}
  \label{fig:tpavsfull_etmiss}
\end{figure}%
Due to the cancellation that is also observed in the integrated cross sections, differential cross sections in observables that are insensitive to non-resonant contributions show the same pattern.
This can be seen for example in the differential distribution of the azimuthal-angle difference of the two positively-charged final-state leptons, $\Delta\phi_{\ell^+_1\ell^+_2}$, depicted in Fig.~\ref{fig:tpavsfull_deltaphi} or in distributions in the pseudorapidity difference $\Delta\eta_{\Pl\ell^+_1}$ of the negatively-charged lepton $\Plm$ and the leading positively-charged lepton $\ell^+_1$, defined via $p_{\rT,\ell^+_1}\geq p_{\rT,\ell^+_2}$, shown in Fig.~\ref{fig:tpavsfull_deltaeta}.
Applying the TPA to the EW virtual corrections cancels the effect of applying the TPA to the QCD virtual corrections.
The \etmiss distribution, on the other hand, is highly sensitive to non-resonant contributions of the form illustrated in Fig.\ \ref{fig:tpavsfull_etmiss_illustration}.
This results in substantial differences of the differential cross section evaluated in TPA or fully off-shell at large missing transverse energies, leading to large values of $\Delta_\TPA$, as can be seen in Fig.~\ref{fig:tpavsfull_etmiss}.
While the QCD corrections in the high-\etmiss\ region are ruled by quark--gluon-induced and gluon-real-emission contributions, which are both always evaluated fully off shell, the EW virtual contributions have a significant impact on the EW corrections in this region.
This results in fairly small values of $\Delta_\TPA^\QCD$, but large values of $\Delta_\TPA^\EW$ in the high-\etmiss\ region.
We observe that the estimate $\Delta_\TPA^\text{estimate}$ generally describes the size of the observed difference $\Delta_\TPA$ between TPA and full off-shell calculation well.

\section{Conclusion}
\label{sec:Conclusion}
We have presented a calculation of hadronic $\PW\PW\PW$ production at the LHC with leptonic \PW-boson decays including NLO EW and QCD corrections.
Using $2\rightarrow6/7$ amplitudes, we have evaluated integrated and differential cross sections taking into account the full off-shell and spin correlation information as well as intermediary resonances.
We observe, similarly to the case of $\PW\PW\PW$ production with stable \PW bosons, a strong but accidental cancellation among the quark--photon and quark--antiquark-induced EW corrections.
For the chosen event setup, they are of similar size ($\sim$\,4--8\,\%) but different in sign, so that the total EW corrections are below the percent level.
QCD corrections at the CM energy of the LHC of $\sqrt{s}=13\,\TeV$ amount to approximately $40\,\%$.
As the analyzed process is independent of $\alphas$ at LO, we do not see a decrease of the residual scale dependence from LO to NLO.
To obtain a reduction of the scale uncertainty, next-to-next-to-leading order (NNLO) QCD calculations or multi-jet merging would be necessary.

In differential distributions we observe a strong impact of the EW high-energy logarithms, which reach 20--30\,\% in the \TeV range.
Angular distributions are slightly modified in shape when including NLO corrections.
Thus, to constrain anomalous gauge couplings, the NLO corrections presented in this paper should be included.

Apart from the full off-shell calculation, we have further performed a calculation within the triple-pole approximation (TPA).
The TPA is based on the leading term in the expansion of the one-loop matrix elements around the resonances of the three \PW bosons.
We have compared results of the TPA with results of the full off-shell calculation in a setup that excludes the Higgs-strahlung subprocess, which can be achieved due to the good separation originating from the small Higgs width and the mass hierarchy $\MH<2\MW$.
The TPA performs very well in integrated cross sections and in angular and rapidity distributions, which are insensitive to off-shell effects.
In this context we observe an accidental cancellation of the TPA error (w.r.t.\ the full off-shell calculation) between NLO EW and QCD corrections.
For some observables, however, that become sensitive to non-resonant contributions, like the missing transverse momentum at high scales, the TPA is not a sufficient approximation.
Sizable deviations can be observed in these regions.
Nevertheless, the size of the TPA uncertainty can be estimated reasonably well to identify those regions by analyzing TPA results only.

In summary, the presented NLO results for EW corrections based on the full off-shell matrix elements are certainly sufficient for the analyses of $\PW\PW\PW$ production at the LHC.
For integrated cross sections, even NLO EW corrections in the TPA will be sufficiently precise.

\section*{Acknowledgments}
The authors thank Marek Schönherr for a detailed comparison of the presented numerical results in integrated cross sections to the ones shown in Ref.~\cite{Schonherr:2018jva}.

S.D.\ and G.K.\ are supported by the Research Training Group GRK 2044 of the German Research Foundation (DFG).
S.D.\ and C.S.\ acknowledge support by the DFG through grant DI 784/3.
Moreover, C.S.\ is supported by the European Research Council under the European Unions Horizon 2020 research and innovation Programme (grant agreement no.\ 740006).
The authors acknowledge support by the state of Baden-Württemberg through bwHPC and the DFG through grant no INST 39/963-1 FUGG.
\FloatBarrier
\bibliographystyle{tep}
\bibliography{www}
\end{document}